\documentclass[prx, twocolumn, superscriptaddress]{revtex4-2}
\usepackage{bm, amsmath, amsfonts, amssymb, mathtools, braket}
\usepackage{times}
\usepackage{multirow}
\usepackage{graphicx}
\usepackage{float, color, xcolor}
\usepackage{booktabs}
\usepackage{bbm}
\usepackage{tabularx}
\usepackage{makecell}
\usepackage[
pagebackref=false,
colorlinks=true,
linkcolor=blue,
urlcolor=blue,
filecolor=black,
citecolor=red,
pdfstartview=FitV,
pdftitle={},
pdfauthor={},
pdfsubject={},
pdfkeywords={},
pdfpagemode=None,
bookmarksopen=true
]{hyperref}

\usepackage[normalem]{ulem}
\usepackage{diagbox}
\newcommand{\tr}{\mathrm{Tr}}
\newcommand{\ra}{\rightarrow}
\newcommand{\bo}{\mathbbm{1}}
\newcommand{\dg}{\dagger}

\newcommand{\sgn}{\mathsf{s}}

\newcommand{\HBCFT}{H_{\mathrm{CFT}}}

\usepackage{tikz}
\usetikzlibrary{arrows, arrows.meta,fadings}
\usetikzlibrary{decorations.pathreplacing,calligraphy,decorations.markings}

\definecolor{tensorcolor}{rgb}{0.65,0.77,0.95}
\definecolor{ttcolor}{rgb}{0.6,0.8,0.2}
\definecolor{tdcolor}{rgb}{0.7,0.9,1.0}
\definecolor{ttcolordark}{rgb}{0.4,0.6,0.1}
\definecolor{tdcolordark}{rgb}{0.5,0.7,0.8}
\definecolor{mpdocolor}{rgb}{1,0.93,0.83}
\definecolor{btensorcolor}{rgb}{0.65,0.50,0.69}
\definecolor{whitetensorcolor}{rgb}{0.99,0.99,0.99}
\definecolor{diamondcolor}{rgb}{0.65,0.77,0.95}
\definecolor{cutoffcolor}{rgb}{0.3,0.4,0.9}
\definecolor{unitarycolor}{rgb}{0.8,0.5,.5}
\definecolor{lcolor}{rgb}{0.94,0.97,1.}

\definecolor{boxc1}{rgb}{0.13,0.54,0.13}
\definecolor{boxc1l}{rgb}{0.69, 0.83, 0.68}
\definecolor{boxc2}{rgb}{0.37,0.62,0.63}
\definecolor{boxc3}{rgb}{0.11,0.56,1.}
\definecolor{boxc4}{rgb}{0.5,0.5,0.5}
\definecolor{boxc5}{rgb}{0.3,0.3,0.3}

\newcommand{\mthick}{thick}

\newcommand{\GcTensor}[7]{
    \begin{scope}[shift={(#1)}]
    \ifnum#5=0
		\draw[\mthick] (-#2,0) -- (#2,0);
		\draw[\mthick] (0,#2) -- (0,-#2);
    \fi
    \ifnum#5=-1
		\draw[\mthick] (0,0) -- (-#2,0);
		\draw[\mthick] (0,0) -- (0,#2);
    \fi
    \ifnum#5=1
		\draw[\mthick] (#2,0) -- (0,0);
		\draw[\mthick] (0,0) -- (0,#2);
    \fi

    \ifnum#5=2
		\draw[\mthick] (-#2,0) -- (#2,0);
		\draw[\mthick] (0,0) -- (0,#2);
    \fi
    \ifnum#5=-2
		\draw[\mthick] (-#2,0) -- (#2,0);
		\draw[\mthick] (0,0) -- (0,-#2);
    \fi

    \ifnum#5=3
		\draw[\mthick] (-#2,0) -- (#2,0);
		\draw[\mthick] (0,-#2) -- (0,#2);
    \fi
    \ifnum#5=-4
		\draw[\mthick] (0,0) -- (-#2,0);
		\draw[\mthick] (0,0) -- (0,-#2);
    \fi
    \ifnum#5=4
		\draw[\mthick] (#2,0) -- (0,0);
		\draw[\mthick] (0,0) -- (0,-#2);
    \fi
    \ifnum#5=5
		\draw[\mthick] (-#2,0) -- (#2,0);
    \fi
        \draw[ thick, fill=#6, rounded corners=2pt] (-#3,-#3) rectangle (#3,#3);
		\draw (0,#7) node {\scriptsize #4};
	\end{scope}
}

\newcommand{\bigtensor}[3]{
\begin{scope}[shift={(#1)}]
    \ifnum#2=1
        \foreach \x in {1,...,#3}{
            \draw[\mthick] (\singledx*\x,0) -- (\singledx*\x,1.2);
        }
    \fi
    \ifnum#2=2
        \foreach \x in {1,...,#3}{
            \draw[\mthick] (\singledx*\x,0) -- (\singledx*\x,-1.2);
        }
    \fi
    \draw[\mthick, fill=white, rounded corners=2pt] (\singledx-0.5,-0.5) rectangle (\singledx*#3+0.5,0.5);
\end{scope}
}

\newcommand{\GtTensor}[6]{
\ifnum#6=2
		\GcTensor{#1}{#2}{#3}{#4}{#5}{ttcolor}{-#2*1.1};
    \fi
\ifnum#6=1
		\GcTensor{#1}{#2}{#3}{#4}{#5}{tdcolor}{-#2*1.1};
    \fi
\ifnum#6=3
		\GcTensor{#1}{#2}{#3}{#4}{#5}{tensorcolor}{-#2*1.1};
    \fi
\ifnum#6=4
		\GcTensor{#1}{#2}{#3}{#4}{#5}{tdcolordark}{0,0};
    \fi
\ifnum#6=5
		\GcTensor{#1}{#2}{#3}{#4}{#5}{ttcolordark}{0,0};
    \fi
\ifnum#6=0
		\GcTensor{#1}{#2}{#3}{#4}{#5}{white}{-#2*1.1};
    \fi
}

\newcommand{\MPOTensor}[5]{
\begin{scope}[shift={(#1)}]
    \ifnum#5=0
		\draw[\mthick] (-#2,0) -- (#2,0);
		\draw[\mthick] (0,#2) -- (0,-#2);
    \fi
    \ifnum#5=1
		\draw[\mthick] (0,0) -- (#2,0);
		\draw[\mthick] (0,#2) -- (0,-#2);
    \fi
    \ifnum#5=-1
		\draw[\mthick] (-#2,0) -- (0,0);
		\draw[\mthick] (0,#2) -- (0,-#2);
    \fi
        
  \filldraw[color=black, fill=whitetensorcolor, thick] (0,0) circle (#3);
	\draw (#2*0.1,-#2*1.3) node {#4};
\end{scope}
}

\newcommand{\zslicenew}[1]{
\begin{scope}[shift={(#1)}]
\def\ep{0.25}
\def\sf{0.08}
    \def\L{4}
        \draw[very thick] (0,-2.5)--(0,0);
        \draw[very thick] (\L,-2.5)--(\L,0);
        \draw[very thick] (0,2.5)--(0,0);
        \draw[very thick] (\L,2.5)--(\L,0);
        \draw[thick, gray] (0,-\sf) -- (\L*0.5,-\sf);
        \draw[thick, gray] (0,\sf) -- (\L*0.5,\sf);
        \filldraw[color=cutoffcolor, fill=white, thick] (0,0) circle (\ep);
        \draw[ thick, white, fill=white] (-\ep,-\ep) rectangle (-\sf,\ep);
        \filldraw[color=cutoffcolor, fill=white, thick] (\L,0) circle (\ep);
        \draw[ thick, white, fill=white] (\L+\sf,-\ep) rectangle (\L+\ep,\ep);
        \filldraw[color=diamondcolor, fill=white, thick] (\L*0.5,0) circle (\ep);
        \draw[very thick, white] (0,-\ep*0.0) -- (\L*0.5,-\ep*0.0);
        \draw[>=stealth, <->] (0+0.1,-2+0.2) -- (\L-0.1,-2+0.2);
        \draw(\L*0.5,-2-0.3) node {\small $L$};
        \draw[>=stealth, <->] (0+0.1,0.4) -- (\L*0.5-0.1,0.4);
        \draw(\L*0.25,0.8) node {\scriptsize $L_A$};
        \draw(\L*0.25,-0.6) node {\scriptsize $A$};
\end{scope}
}

\newcommand{\wslicenew}[3]{
\begin{scope}[shift={(#1)}]
    \def\ep{0.25}
    \def\L{4}
    \def\y{1.2}
        \draw[thick, color=diamondcolor] (0,-\y)--(0,\y);
        \draw[thick, color=cutoffcolor] (\L,-\y)--(\L,\y);
        \ifnum#3=1
            \draw[thick, color=gray](0,\y) -- (\L,\y);
            \draw[thick, color=gray](0,-\y) -- (\L,-\y);
        \fi
        \ifnum#3=2
            \draw[thick, color=gray](0,0) -- (\L,0);
            \draw[thick, color=gray, densely dotted](0,-\y) -- (\L,-\y);
        \fi
        \ifnum#2=1
            \draw[>=stealth, <->] (0+0.01,-\y-0.4) -- (\L-0.01,-\y-0.4);
            \draw(\L*0.5,-\y-1) node {\small $W$};
            \draw(\L*0.5,-\y+0.5) node {\scriptsize $A$};
            \draw[>=stealth, <->] (\L+0.4,\y) -- (\L+0.4,-\y);
            \draw(\L+1,0) node {\small $2\pi$};
        \fi
\end{scope}
}

\newcommand{\dTensor}[4]{
    \begin{scope}[shift={(#1)}]
        \draw[\mthick] (-#2,0) -- (#2,0);
        \draw[thick, fill=white] (#3,0) -- (0,#3) -- (-#3,0) -- (0,-#3) -- cycle;
        \draw (0,0) node {\scriptsize #4};
    \end{scope}
}

\newcommand{\projector}[7]{
    \begin{scope}[shift={(#1)}]
    \def\ro{.6};
    \def\ls{1.2};
    \def\rs{0.8};
    \def\offset{0.08};
    \ifnum#7=0
		\draw[very thick](-#4*\ls,-#5) -- (-#4*\ls,#5*3);
        \draw[gray, very thick](-#4*\rs,-#5) -- (-#4*\rs,#5*3);
        \draw[gray, very thick](#4*\ls,-#5) -- (#4*\ls,#5-\offset*1.5);
        \draw[very thick, rounded corners=2pt](#4*\rs,-#5) -- (#4*\rs,#5)--(#4*\ls,#5)--(#4*\ls,#5-\offset);
    \fi
    \ifnum#7=1
        \draw[very thick](#4*\rs,-#5) -- (#4*\rs,#5*3);
        \draw[gray, very thick](-#4*\rs,-#5) -- (-#4*\rs,#5-1.5*\offset);
        \draw[gray, very thick](#4*\ls,-#5) -- (#4*\ls,#5*3);
        \draw[very thick, rounded corners=2pt](-#4*\ls,-#5)--(-#4*\ls,#5)--(-#4*\rs,#5)--(-#4*\rs,#5-\offset);
    \fi
    \ifnum#7=2
        \draw[gray, very thick](-#4*\rs,-#5) -- (-#4*\rs,#5-\offset*1.5);
        \draw[gray, very thick](#4*\ls,-#5) -- (#4*\ls,#5-\offset*1.5);
        \draw[very thick, rounded corners=2pt](-#4*\ls,-#5)--(-#4*\ls,#5)--(-#4*\rs,#5)--(-#4*\rs,#5-\offset);
        \draw[very thick, rounded corners=2pt](#4*\rs,-#5) -- (#4*\rs,#5)--(#4*\ls,#5)--(#4*\ls,#5-\offset);
    \fi
        \draw[thick, fill=whitetensorcolor, rounded corners=2pt] (-#2,-#3) rectangle (#2, #3);
        \draw (0,0) node {\scriptsize #6};
    \end{scope}
}

\newcommand{\SingleDots}[2]{
	\begin{scope}[shift={(#1)}]
      \draw [very thick, dotted] (-#2/2,0) to (#2/2,0);
	\end{scope}
}

\newcommand{\SideIdentityTensor}[4]{
	\begin{scope}[shift={(#1)}]
    \ifnum#4=-1
	   \draw [very thick] (\doubledx-1,0.8) to  [bend right=90] (\doubledx-1,-0.8);
    \fi
    \ifnum#4=-2
	   \draw [very thick] (\doubledx-1,0.8) to  [bend right=90] (\doubledx-1,-0.8);
      \draw [very thick] (\doubledx-1,0.8) -- (\doubledx-0.5,0.8);
      \draw [very thick] (\doubledx-1,-0.8) -- (\doubledx-0.5,-0.8);
    \fi
    \ifnum#4=-3
	   \draw [very thick] (\doubledx-1,0.8) to  [bend right=90] (\doubledx-1,-0.8);
      \draw [very thick] (\doubledx-1,0.8) -- (\doubledx-0.5,0.8);
      \draw [very thick] (\doubledx-1,-0.8) -- (\doubledx-0.5,-0.8);
	\filldraw[color=black, fill=whitetensorcolor, thick] (\doubledx-1.4,0) circle (#3);
	\draw (\doubledx-1.4,0) node {#2};
    \fi
    \ifnum#4=1
	   \draw [very thick] (-\doubledx+1,0.8) to  [bend left=90] (-\doubledx+1,-0.8);
    \fi
    \ifnum#4=2
	   \draw [very thick] (-\doubledx+1,0.8) to  [bend left=90] (-\doubledx+1,-0.8);
      \draw [very thick] (-\doubledx+1,0.8) -- (-\doubledx+0.5,0.8);
      \draw [very thick] (-\doubledx+1,-0.8) -- (-\doubledx+0.5,-0.8);
    \fi
    \ifnum#4=3
	   \draw [very thick] (-\doubledx+1,0.8) to  [bend left=90] (-\doubledx+1,-0.8);
      \draw [very thick] (-\doubledx+1,0.8) -- (-\doubledx+0.5,0.8);
      \draw [very thick] (-\doubledx+1,-0.8) -- (-\doubledx+0.5,-0.8);
	\filldraw[color=black, fill=whitetensorcolor, thick] (-\doubledx+1.4,0) circle (#3);
	\draw (-\doubledx+1.4,0) node {#2};
    \fi
\end{scope}
}

\newcommand\doubledx{1.6}
\newcommand\singledx{1.8}

\newcommand{\rd}{0.5}

\definecolor{DB}{rgb}{0.36, 0.54, 0.66}

\definecolor{MDG}{rgb}{0,0.55,0.05}

\definecolor{yuhancolor}{rgb}{0.9, 0, 0.5}

\begin{document}

\title{Excited states from local effective Hamiltonians of matrix product states \\ and their entanglement spectrum transition}

\author{Denise Cocchiarella}
\affiliation{Max-Planck-Institut f{\"u}r Quantenoptik, Hans-Kopfermann-Str. 1, D-85748 Garching, Germany}
\affiliation{Munich Center for Quantum Science and Technology (MCQST), Schellingstr. 4, D-80799 M{\"u}nchen, Germany}

\author{Mingru Yang}
\affiliation{Max-Planck-Institut f{\"u}r Quantenoptik, Hans-Kopfermann-Str. 1, D-85748 Garching, Germany}
\affiliation{Munich Center for Quantum Science and Technology (MCQST), Schellingstr. 4, D-80799 M{\"u}nchen, Germany}

\author{Yueshui Zhang}
\affiliation{Faculty of Physics and Arnold Sommerfeld Center for Theoretical Physics,
Ludwig-Maximilians-Universität München, 80333 Munich, Germany}

\author{Mari Carmen Ba\~nuls}
\affiliation{Max-Planck-Institut f{\"u}r Quantenoptik, Hans-Kopfermann-Str. 1, D-85748 Garching, Germany}
\affiliation{Munich Center for Quantum Science and Technology (MCQST), Schellingstr. 4, D-80799 M{\"u}nchen, Germany}

\author{Hong-Hao Tu}
\affiliation{Faculty of Physics and Arnold Sommerfeld Center for Theoretical Physics,
Ludwig-Maximilians-Universität München, 80333 Munich, Germany}

\author{Yuhan Liu}
\email{yuhan.liu@mpq.mpg.de}
\affiliation{Max-Planck-Institut f{\"u}r Quantenoptik, Hans-Kopfermann-Str. 1, D-85748 Garching, Germany}
\affiliation{Munich Center for Quantum Science and Technology (MCQST), Schellingstr. 4, D-80799 M{\"u}nchen, Germany}

\date{\today}

\begin{abstract}
Solving excited states is a challenging task for interacting systems. For one-dimensional critical systems, however, excited states can be directly accessed from the eigenvectors of the local effective Hamiltonian that is constructed from the ground state obtained by variational matrix product state (MPS) optimization. Despite its numerical success, the theoretical mechanism underlying this method has remained largely unexplored. In this work, we provide a conformal field theory (CFT) perspective that helps elucidate this connection. The key insight is that this construction effectively uses a truncated basis of ground-state Schmidt vectors to represent excited states, where the contribution of each Schmidt vector can be expressed as a CFT correlation function and shown to decay with increasing Schmidt index. The CFT analysis further predicts an entanglement-spectrum transition of excited states as the ratio of the subsystem size to the total system size is varied. Our numerical results support this picture and demonstrate a reorganization of the entanglement spectrum into distinct conformal towers as this ratio changes. 
\end{abstract}

\maketitle

\section{Introduction}

Understanding interacting quantum many-body systems remains a central challenge in modern physics. 
The density-matrix renormalization group (DMRG)~\cite{White1992} has become a cornerstone numerical method for tackling such systems, particularly in one dimension. Its success is intimately connected to the tensor network representation of quantum states, notably the matrix product state (MPS) ansatz~\cite{Ostlund1995,Verstraete2004a,PerezGarcia2006,Schollwoeck2011,Cirac2021}, which captures the entanglement features of quantum states and provides both a conceptual and computational framework for DMRG and various generalizations~\cite{Verstraete2008,Cirac2009,Orus2014,Silvi2019,Okunishi2022,Banuls2023,XiangT-Book}.

In one-dimensional (1D) systems, a particularly important class is that of critical systems, which sit at quantum phase transition points. Their ground-state correlation functions and entanglement properties exhibit universal behaviors, largely independent of microscopic details and described by conformal field theory (CFT)~\cite{Cardy1984,Holzhey1994,Vidal2003a,Calabrese2004,HuQ2020,LiuYH2024}.
For finite systems, 
MPS can efficiently approximate such ground states~\cite{Verstraete2006}, and the DMRG algorithm is successfully used to study critical properties.
Remarkably, even though DMRG is primarily designed for ground states, it can also provide accurate information about excited states in critical systems. As observed in~\cite{Chepiga2017}, the excitation spectrum of the local effective Hamiltonian produced during DMRG sweeps accurately reproduces the low-energy excitations of the original model, enabling explicit construction of excited states. Despite the proven numerical success and growing applications of this approach~\cite{Eberharter2023,LiX2024,Cocchiarella2025,Kovalska2025}, its theoretical foundation remains largely unexplored.

In this paper, we provide a CFT perspective to elucidate the theoretical mechanism behind this numerical observation. We show that, when a 1D critical chain is bipartitioned near its center, the truncated basis formed by the first $D$ Schmidt vectors of the ground state can already capture the essential features of certain low-lying excited states $a$ (that correspond to the primary states of CFT). To quantify this, we introduce a measure $R_a(D)$ that characterizes the accuracy of this truncated basis and relate the contribution of each Schmidt vector to a correlation function in the CFT. The CFT analysis shows that this correlation function decays with increasing Schmidt index, therefore providing qualitative insight into why a finite number of ground-state Schmidt vectors suffices to represent low-lying excited states.

As a byproduct, the same correlation function also encodes information about the reduced density operator of the excited state, from which the entanglement spectrum can be obtained. The CFT analysis suggests that varying the ratio $r$ between the subsystem size and the total chain length may modify the reduced density matrix, thereby inducing a transition in the entanglement spectrum. We support this picture by numerical calculations: at $r=1/2$, the entanglement spectrum of excited states still organizes into conformal towers, although different from those in the ground-state entanglement spectrum. This structure contrasts sharply with the spectrum in the opposite limit $r\ll 1$~\cite{Alcaraz2011,IbanezBerganza2012}, signaling a transition in the entanglement spectrum as $r$ increases from a small value to $1/2$. 

This paper is organized as follows. In Sec.~\ref{sec:excited-state}, we describe how low-lying excited states can be obtained using a truncated basis of ground-state Schmidt vectors together with the local effective Hamiltonian, and introduce measures to quantify the accuracy of this approach. In Sec.~\ref{sec:cft}, we employ conformal field theory to explain why this truncated basis provides an accurate description for excited states that correspond to primary states of the CFT. The CFT analysis further predicts an entanglement-spectrum transition, which we confirm numerically in Sec.~\ref{sec:transition}. We conclude in Sec.~\ref{sec:discussion} with a summary and outlook on open questions. Appendix~\ref{app:Gaussian} contains technical details on the free-fermion methods used to compute entanglement quantities in the transverse-field Ising chain.

\section{Excited states from local effective Hamiltonian}
\label{sec:excited-state}

In this section, we outline the procedure for obtaining low-lying excited states in 1D critical systems through the local effective Hamiltonian. Consider a 1D quantum system consisting of $N$ spins with local Hilbert space dimension $d$, so that the full Hilbert space is $\mathcal{H} =(\mathbb{C}^{d})^{\otimes N}$. A Hamiltonian is a Hermitian operator acting on $\mathcal{H}$, denoted by $H$. Throughout this work, we focus on local Hamiltonians of the form $H=\sum_i h_i$, where each local term $h_i$ has finite spatial support.

The DMRG algorithm is a variational method which approximates the ground state of $H$ as an open-boundary matrix product state (MPS)~\cite{Ostlund1995,Verstraete2004a,Verstraete2008}, graphically represented as
\begin{align}
|\psi_0\rangle = 
    \begin{array}{c}
        \begin{tikzpicture}[scale=.4,baseline={([yshift=-0.75ex] current bounding box.center)}]
		      \foreach \x in {1,...,1}{
                \GtTensor{(-\singledx*\x,0)}{1}{\rd}{\small $B^N$}{-1}{0}
        }
		      \foreach \x in {2,...,2}{
                \SingleDots{-\singledx*\x,0}{\singledx/2}
        }
                \foreach \x in {3,...,3}{
                \GtTensor{(-\singledx*\x,0)}{1}{\rd}{\small $B^3$}{2}{0}
        }
		      \foreach \x in {4,...,4}{
                \GtTensor{(-\singledx*\x,0)}{1}{\rd}{\small $B^2$}{2}{0}
        }
        \foreach \x in {5,...,5}{
                \GtTensor{(-\singledx*\x,0)}{1}{\rd}{\small $B^1$}{1}{0}
        }
        \end{tikzpicture}
        \end{array}
        \,.
\end{align}
The state is parametrized by $N$ local tensors $\{ B^i\}_{i=1,2,\ldots,N}$. Each $B^i$ is a rank-3 tensor with one physical leg (vertical bond) of dimension $d$ and two virtual legs (left and right horizontal bonds) of dimension $D_i$ and $D_{i+1}$. The boundary tensors $B^1$ and $B^N$ have $D_1 = D_{N+1} = 1$. We define $D$ as the largest virtual bond dimension, $D=\max\{D_i\}$. If $H$ is gapped, its ground state satisfies an area law of entanglement~\cite{Eisert2010} and can be efficiently approximated by an MPS with finite bond dimension~\cite{Hastings2007}. For a 1D critical Hamiltonian described by a CFT, the ground-state entanglement entropy between two halves of the system grows logarithmically with the system size~\cite{Holzhey1994,Vidal2003a,Calabrese2004}.
Yet, the ground state can be approximated by an MPS whose bond dimension scales polynomially with the system size, which ensures that MPSs can still provide a good approximation to the ground state of 1D critical systems of finite size $N$.

Given the ground state $|\psi_0\rangle$ in MPS form, we divide the chain into a left subsystem $A$ consisting of sites $1$ to $N_A$ ($N_A \leq N/2$) and its complement $\bar{A}$ comprising sites $N_A + 1$ to $N$. The MPS can then be transformed into a canonical form associated with this bipartition:
\begin{align}
|\psi_0\rangle = 
    \begin{array}{c}
        \begin{tikzpicture}[scale=.4,baseline={([yshift=-0.75ex] current bounding box.center)}]
		      \foreach \x in {1,...,1}{
                \GtTensor{(-\singledx*\x,0)}{1}{\rd}{\small $B^N$}{-1}{2}
        }
                \foreach \x in {2,...,2}{
                \GtTensor{(-\singledx*\x,0)}{1}{\rd}{\small $B^{N-1}$}{2}{2}
        }
		      \foreach \x in {3,...,3}{
                \SingleDots{-\singledx*\x,0}{\singledx/2}
        }
                \foreach \x in {4,...,4}{
                \GtTensor{(-\singledx*\x,0)}{1}{\rd}{\small $B^{N_A+1}$}{2}{2}
        }
		      \foreach \x in {5,...,5}{
                \dTensor{(-\singledx*\x,0)}{1}{0.6}{\small $\Lambda_0$}
        }
                \foreach \x in {6,...,6}{
                \GtTensor{(-\singledx*\x,0)}{1}{\rd}{\small $B^{N_A}$}{2}{1}
        }
        \foreach \x in {7,...,7}{
                \SingleDots{-\singledx*\x,0}{\singledx/2}
        }
        \foreach \x in {8,...,8}{
                \GtTensor{(-\singledx*\x,0)}{1}{\rd}{\small $B^2$}{2}{1}
        }
        \foreach \x in {9,...,9}{
                \GtTensor{(-\singledx*\x,0)}{1}{\rd}{\small $B^1$}{1}{1}
        }
        \end{tikzpicture}
        \end{array}
        \,,
\end{align}
where MPS tensors at sites $1,2,\ldots,N_A$ ($N_A+1,\ldots,N$) satisfy the left (right) canonical condition,
\begin{equation}
    \begin{array}{c}
        \begin{tikzpicture}[scale=.4,baseline={([yshift=5ex] current bounding box.center)}]
        \draw[\mthick, rounded corners=2pt](0,0) -- (-\singledx*0.75,0) -- (-\singledx*0.75, \singledx) -- (0, \singledx); 
                \GtTensor{(0,0)}{1}{\rd}{}{2}{1}
                \GtTensor{(0,\singledx)}{1}{\rd}{}{-2}{1}
        \end{tikzpicture}
        \end{array}=
        \begin{array}{c}
        \begin{tikzpicture}[scale=.4,baseline={([yshift=-0ex] current bounding box.center)}]
        \draw[\mthick, rounded corners=2pt](0,0) -- (-\singledx*0.5,0) -- (-\singledx*0.5, \singledx) -- (0, \singledx); 
        \end{tikzpicture}
        \end{array},\quad \quad 
         \begin{array}{c}
        \begin{tikzpicture}[scale=.4,baseline={([yshift=5ex] current bounding box.center)}]
        \draw[\mthick, rounded corners=2pt](0,0) -- (\singledx*0.75,0) -- (\singledx*0.75, \singledx) -- (0, \singledx); 
                \GtTensor{(0,0)}{1}{\rd}{}{2}{2}
                \GtTensor{(0,\singledx)}{1}{\rd}{}{-2}{2}
        \end{tikzpicture}
        \end{array}=
        \begin{array}{c}
        \begin{tikzpicture}[scale=.4,baseline={([yshift=0ex] current bounding box.center)}]
        \draw[\mthick, rounded corners=2pt](0,0) -- (\singledx*0.5,0) -- (\singledx*0.5, \singledx) -- (0, \singledx); 
        \end{tikzpicture}
        \end{array}
        \,,
\end{equation}
and $\Lambda_0$ is a $D \times D$ diagonal matrix with descendingly ordered diagonal entries $\lambda_\alpha > 0$ $(\alpha=0,\ldots,D-1)$, normalized as $\sum_{\alpha=0}^{D-1} \lambda_\alpha^2=1$. This bond canonical form can always be achieved by successive singular value decompositions~\cite{Vidal2003b,PerezGarcia2006}. It yields the Schmidt decomposition of the MPS with respect to the $A|\bar{A}$ bipartition: 
\begin{equation}
    |\psi_0\rangle = \sum_{\alpha=0}^{D-1}\lambda_\alpha |v_\alpha\rangle\otimes|w_\alpha\rangle,
    \label{eqn:Schmidt-MPS}
\end{equation}
where $|v_\alpha\rangle$ and $|w_\beta\rangle$ are the Schmidt vectors of two subsystems:
\begin{equation}
\begin{aligned}
    |v_\alpha\rangle&=\begin{array}{c}
        \begin{tikzpicture}[scale=.4,baseline={([yshift=-0.75ex] current bounding box.center)}]
		      \foreach \x in {1,...,1}{
                \GtTensor{(\singledx*\x,-\singledx)}{1}{\rd}{}{1}{1}
        }
		      \foreach \x in {2,...,3}{
                 \GtTensor{(\singledx*\x,-\singledx)}{1}{\rd}{}{2}{1}
        }
        \foreach \x in {4,...,4}{
                \GtTensor{(\singledx*\x,-\singledx)}{1}{\rd}{}{2}{1};
        }
        \draw (\singledx*5,-\singledx) node {\small $\alpha$};
        \end{tikzpicture}
        \end{array}\\
    |w_\beta\rangle &=
    \begin{array}{c}
    \begin{tikzpicture}[scale=.4,baseline={([yshift=-0.75ex] current bounding box.center)}]
        \draw (\singledx*0,-\singledx) node {\small $\beta$};
        \foreach \x in {1,...,1}{
        \GtTensor{(\singledx*\x,-\singledx)}{1}{\rd}{}{2}{2};
        }
        \foreach \x in {2,...,3}{
                \GtTensor{(\singledx*\x,-\singledx)}{1}{\rd}{}{2}{2}
        }
        \foreach \x in {4,...,4}{
                \GtTensor{(\singledx*\x,-\singledx)}{1}{\rd}{}{-1}{2}
        }
    \end{tikzpicture}
        \end{array}
\end{aligned}
\end{equation}
with $\alpha,\beta=0,\ldots,D-1$.
Using the left and right canonical conditions, it is easy to check that these vectors indeed satisfy the orthonormality conditions $\langle v_\alpha|v_{\alpha'}\rangle=\delta_{\alpha\alpha'}$ and $\langle w_\beta|w_{\beta'}\rangle=\delta_{\beta\beta'}$.

Local Hamiltonians in one spatial dimension admit an efficient tensor network representation as a Matrix Product Operator (MPO)~\cite{McCulloch2007}, graphically represented as 
\begin{align}
H = 
    \begin{array}{c}
        \begin{tikzpicture}[scale=.45,baseline={([yshift=-0.75ex] current bounding box.center)}]
		      \foreach \x in {1,...,1}{
                \MPOTensor{(-\singledx*\x,0)}{1}{\rd}{\scriptsize $M^N$}{-1}
        }
		      \foreach \x in {2,...,2}{
                \SingleDots{-\singledx*\x,0}{\singledx/2}
        }
                \foreach \x in {3,...,3}{
                \MPOTensor{(-\singledx*\x,0)}{1}{\rd}{\scriptsize $M^3$}{0}
        }
		      \foreach \x in {4,...,4}{
                \MPOTensor{(-\singledx*\x,0)}{1}{\rd}{\scriptsize $M^2$}{0}
        }
        \foreach \x in {5,...,5}{
                \MPOTensor{(-\singledx*\x,0)}{1}{\rd}{\scriptsize $M^1$}{1}
        }
        \end{tikzpicture}
        \end{array}
        \,,
\end{align}
which is a contraction of $N$ local tensors $\lbrace M^i\rbrace_{i=1,2,\ldots,N}$. Using the ground-state Schmidt vectors and the MPO representation of $H$, we can define a local effective Hamiltonian for the $N_A$-th bond, 
constructed as 
\begin{equation}
    H_{\mathrm{eff}} =
    \begin{array}{c}
    \def\sft{0.5}
        \begin{tikzpicture}[scale=.4,baseline={([yshift=7ex] current bounding box.center)}]
        \draw[thick] (\singledx*4,0) -- (\singledx*5+\sft,0);
		      \foreach \x in {1,...,1}{
                \MPOTensor{(\singledx*\x,0)}{1}{\rd}{}{1}
                \GtTensor{(\singledx*\x,-\singledx)}{1}{\rd}{}{1}{1}
                \GtTensor{(\singledx*\x,\singledx)}{1}{\rd}{}{4}{1}
        }
		      \foreach \x in {2,...,3}{
                \MPOTensor{(\singledx*\x,0)}{1}{\rd}{}{0}
                 \GtTensor{(\singledx*\x,-\singledx)}{1}{\rd}{}{2}{1}
                \GtTensor{(\singledx*\x,\singledx)}{1}{\rd}{}{-2}{1}
        }
		      \foreach \x in {4,...,4}{
                \MPOTensor{(\singledx*\x,0)}{1}{\rd}{}{0}
                \GtTensor{(\singledx*\x,-\singledx)}{0.8}{\rd}{}{2}{1};
                \GtTensor{(\singledx*\x,\singledx)}{0.8}{\rd}{}{-2}{1};
        }
        \foreach \x in {5,...,5}{
                \MPOTensor{(\singledx*\x+\sft,0)}{1}{\rd}{}{0}
                \GtTensor{(\singledx*\x+\sft,-\singledx)}{0.8}{\rd}{}{2}{2};
                 \GtTensor{(\singledx*\x+\sft,\singledx)}{0.8}{\rd}{}{-2}{2};
        }
        \foreach \x in {6,...,7}{
                \MPOTensor{(\singledx*\x+\sft,0)}{1}{\rd}{}{0}
                \GtTensor{(\singledx*\x+\sft,-\singledx)}{1}{\rd}{}{2}{2}
                \GtTensor{(\singledx*\x+\sft,\singledx)}{1}{\rd}{}{-2}{2}
        }
        \foreach \x in {8,...,8}{
                \MPOTensor{(\singledx*\x+\sft,0)}{1}{\rd}{}{-1}
                \GtTensor{(\singledx*\x+\sft,-\singledx)}{1}{\rd}{}{-1}{2}
                \GtTensor{(\singledx*\x+\sft,\singledx)}{1}{\rd}{}{-4}{2}
        }
        \end{tikzpicture}
        \end{array}
        \,.
\label{eqn:Heff}
\end{equation}
$H_{\mathrm{eff}}$ is a $D^2\times D^2$ matrix, 
and its lowest eigenvector (assuming it is non-degenerate) corresponds to the vectorized $\Lambda_0$, 
$H_{\mathrm{eff}} \Lambda_0 = E_0 \Lambda_0$, where $E_0$ is the variational ground-state energy. We note that, unlike in standard DMRG algorithms where the effective Hamiltonian is defined on one or two physical sites, $H_{\mathrm{eff}}$ in Eq.~\eqref{eqn:Heff} is defined on the virtual bond. This choice allows for a simpler analysis while still capturing the essential physics. Upon diagonalization, $H_{\mathrm{eff}}$ produces not only $\Lambda_{0}$ but also eigenvectors $\Lambda_a$ ($a=1,\ldots,D^2-1$) corresponding to higher eigenvalues, $H_{\mathrm{eff}} \Lambda_a = E_a \Lambda_a$ with $E_0 < E_1 \leq \cdots \leq E_{D^2-1}$. Each $\Lambda_a$ can be regarded as a $D\times D$ square matrix. As their vectorized forms are eigenvectors of the Hermitian matrix $H_{\mathrm{eff}}$, they can be chosen orthonormal with respect to the Hilbert-Schmidt inner product, $\tr(\Lambda_a^{\dg} \Lambda_{b})=\delta_{ab}$ for $a,b=0,\ldots,D^2-1$.

Using $\Lambda_a$, one can build ``stitched'' excitation Ans\"atze $|\psi_{a}\rangle$ for the higher eigenvectors of the full Hamiltonian $H$:
\begin{equation}
\begin{aligned}
     \Lambda_a &=
    \begin{array}{c}
    \begin{tikzpicture}[scale=.4,baseline={([yshift=-0.75ex] current bounding box.center)}]
             \GcTensor{(0,0)}{1}{\rd}{\small $\Lambda_a$}{5}{whitetensorcolor}{0}
    \end{tikzpicture}
    \end{array}\\
    |\psi_{a}\rangle &=
    \begin{array}{c}
    \def\sft{0.2}
        \begin{tikzpicture}[scale=.4,baseline={([yshift=-0.75ex] current bounding box.center)}]
		      \foreach \x in {1,...,1}{
                \GtTensor{(\singledx*\x,-\singledx)}{1}{\rd}{}{1}{1}
        }
		      \foreach \x in {2,...,3}{
                 \GtTensor{(\singledx*\x,-\singledx)}{1}{\rd}{}{2}{1}
        }
        \foreach \x in {4,...,4}{
                \GtTensor{(\singledx*\x,-\singledx)}{0.8}{\rd}{}{2}{1};
        }
        \GcTensor{(\singledx*5-\sft,-\singledx)}{0.8}{\rd}{\small $\Lambda_a$}{5}{whitetensorcolor}{0};
        \foreach \x in {6,...,6}{
        \GtTensor{(\singledx*\x-\sft*2,-\singledx)}{0.8}{\rd}{}{2}{2};
        }
        \foreach \x in {7,...,8}{
                \GtTensor{(\singledx*\x-\sft*2,-\singledx)}{1}{\rd}{}{2}{2}
        }
        \foreach \x in {9,...,9}{
                \GtTensor{(\singledx*\x-\sft*2,-\singledx)}{1}{\rd}{}{-1}{2}
        }
        \end{tikzpicture}
        \end{array}.
\end{aligned}
\label{eqn:psi0eff}
\end{equation}
These states
are superpositions of products of the ground-state Schmidt vectors
\begin{equation}
    |\psi_a\rangle = \sum_{\alpha,\beta=0}^{D-1} (\Lambda_a)_{\alpha\beta}|v_\alpha\rangle\otimes |w_\beta\rangle,
    \label{eqn:expand0}
\end{equation}
and they form an orthonormal basis, $\langle\psi_a|\psi_b\rangle = \delta_{ab}$. The central question we want to analyze is:  how well the states $\{|\psi_a\rangle\}$ capture the low-lying \textit{physical} excitations of the full Hamiltonian $H$.

To study this question quantitatively, we denote the \textit{exact} eigenstates of the Hamiltonian $H$ by $\{|\Psi_a\rangle\}$, labeled by $a=0,1,\ldots$ in order of increasing energy. The Schmidt decomposition of the exact ground state is written as
\begin{equation}
    |\Psi_0\rangle = \sum_{\alpha=0}^{{d^{N_A}}-1}\tilde{\lambda}_{\alpha} |v_\alpha\rangle\otimes|w_\alpha\rangle
    \label{eqn:Schmidt-GS}
\end{equation}
with Schmidt coefficients arranged in descending order, $\tilde{\lambda}_0 \geq \tilde{\lambda}_1 \geq \cdots \geq 0$, and normalized as $\sum_{\alpha=0}^{d^{N_A}-1} \tilde{\lambda}_{\alpha}^2 = 1$. Roughly speaking, the Schmidt decomposition of the MPS in Eq.~\eqref{eqn:Schmidt-MPS} is a low-rank approximation of Eq.~\eqref{eqn:Schmidt-GS} and the Schmidt vectors in Eq.~\eqref{eqn:Schmidt-MPS} are the same as the first $D$ ones in Eq.~\eqref{eqn:Schmidt-GS} (this is rigorous if no truncations are incurred at other bonds of the MPS). In other words, choosing a bond dimension $D$ amounts to truncating the basis of Schmidt vectors from $d^{N_A}$ vectors to $D$ vectors. 

In general, the exact excited states of $H$ can be expanded in terms of the ground-state Schmidt vectors $\{|v_\alpha\rangle\otimes|w_\beta\rangle\}$ (in particular, the Schmidt vectors form a complete basis when $N_A = N/2$):
\begin{equation}
    |\Psi_a\rangle=\sum_{\alpha,\beta=0}^{d^{N_A}-1} (\tilde{\Lambda}_a)_{\alpha\beta}|v_\alpha\rangle\otimes |w_\beta\rangle
    \label{eqn:Schmidt-ES}
\end{equation}
with $\tr(\tilde{\Lambda}_a^{\dg} \tilde{\Lambda}_b)=\delta_{ab}$. The question of interest is whether its drastically truncated version ($D \ll d^{N_A}$) in Eq.~\eqref{eqn:expand0} is a good approximation, at least for the low-lying excited states.

In order to quantify the difference between the ansatz $\{|\psi_a\rangle\}$ and the true physical excitations $\{|\Psi_a\rangle\}$, we consider two fidelity quantities: (i) the overlap 
\begin{equation}
    |\langle \psi_{a}|\Psi_b\rangle|,
    \label{eqn:overlap}
\end{equation}
where $|\psi_{a}\rangle$ is the stitched excitation ansatz obtained from effective Hamiltonian [see Eq.~\eqref{eqn:psi0eff}] and $|\Psi_b\rangle$ is the exact excited state;  and (ii) the norm of the projection of $| \Psi_a \rangle$ onto the $D^2$-dimensional subspace generated by the products of $D$ low-lying ground-state Schmidt vectors, 
\begin{equation}
R_a(D) = \sum_{\alpha,\beta=0}^{D-1} |(\langle v_{\alpha}| \otimes \langle w_{\beta}| ) | \Psi_a \rangle|^2 ,
\label{eqn:R_aD}
\end{equation}
where $|v_{\alpha}\rangle$ and $|w_{\beta}\rangle$ are the ground-state Schmidt vectors [see Eq.~\eqref{eqn:Schmidt-GS}]. Note that $R_a(d^{N_A}) = 1$ when $N_A=N/2$, which corresponds to the case of no truncation $D=d^{N_A}$. 
A value of $R_a(D)$ close to 1 would indicate that $D$ low-lying Schmidt vectors of $|\Psi_0\rangle$ are enough to represent $|\Psi_a\rangle$ accurately.

To gain some insight, let us take the transverse-field Ising chain (TFIC) with the open boundary condition as a benchmark example:
\begin{equation}
    H_{\text{Ising}} = -\sum_{i=1}^{N-1} \sigma^z_i \sigma^z_{i+1} - g \sum_{i=1}^N \sigma^x_i ,
    \label{eqn:Ising-Hamil}
\end{equation}
where $\sigma^x$ and $\sigma^z$ are the $2\times 2$ Pauli matrices. The TFIC is critical at $|g|=1$ (described by the Ising CFT) and gapped for $|g| \neq 1$. It is worth noting that the TFIC can be mapped onto a free fermion model~\cite{Pfeuty1970} whose eigenstates are fermionic Gaussian states. This property allows for an exact computation of the Schmidt decomposition, as well as the fidelity quantity $R_a(D)$ discussed above. Further technical details are summarized in Appendix~\ref{app:Gaussian}. 

We numerically computed these two fidelity quantities for the TFIC with chain length $N=40$ and truncation dimension $D=30$. As shown in Table~\ref{tab:good-approx}, the state overlaps behave as $|\langle\psi_a|\Psi_b\rangle|\approx \delta_{ab}$ at the critical point $g=1$, with errors below $10^{-5}$ for the ground state and first three excited states. Similarly, Table~\ref{tab:R-table} shows that when $N_A=N/2$, $R_a(D)\approx 1$ at the critical point $g=1$, and the deviation remains within $10^{-5}$ for the lowest four states. In contrast, at the non-critical point $g=1.5$, $R_a(D)$ clearly deviates from unity, indicating a clear loss of accuracy away from criticality.

\begin{table}[]
\centering
\begin{tabular}{c| c c c c}
\toprule \toprule
\diagbox{$a$}{$b$} & 0 & 1 & 2 & 3 \\
\midrule
0 
& $1- 8 \times 10^{-15}$ & $10^{-13}$ & $2 \times 10^{-9}$ &  $4 \times 10^{-9}$\\
1 
& $2 \times 10^{-13}$ & $1- 2 \times 10^{-8}$ & $ 10^{-9}$ &  $8 \times 10^{-9}$\\
2 
&$3 \times 10^{-13}$ & $10^{-12}$ & $1- 2 \times 10^{-6}$ & $10^{-5}$ \\
3 &
$6 \times 10^{-12}$ & $2 \times 10^{-13}$ & $10^{-5}$&  $1-9 \times 10^{-6}$ \\
\bottomrule \bottomrule
\end{tabular}
    \caption{Results of $|\langle \psi_{a}|\Psi_b\rangle|$ for the TFIC at the critical point $g=1$, for the ground state ($a=0$) and first three excited states ($a=1,2,3$). The chain length is $N=40$, and the truncation dimension for the stitched excitation ansatz $|\psi_a\rangle$ (namely, dimension of $\Lambda_a$) is chosen as $D=30$. The (numerically) exact excited state $|\Psi_b\rangle$ is obtained from DMRG with a large bond dimension $\tilde{D}=1000$ using ITensor Software Library~\cite{itensor}.}
    \label{tab:good-approx}
\end{table}


\begin{table}[t]
\centering
\begin{tabular}{c|c c c}
\toprule \toprule
$a$ &
\begin{tabular}{c} $R_a(D)$ \\ $g=1.0$ \end{tabular} &
\begin{tabular}{c} $R_a(D)$ \\ $g=1.5$ \end{tabular} &
\begin{tabular}{c} $R'_a(D)$ \\ $g=1.0$ \end{tabular} \\
\midrule
0 & $1-2\times 10^{-14}$ & $1-5\times 10^{-14}$ &$1-1\times 10^{-14}$\\
1 &  $1-2\times 10^{-8}$ & 0.9226 & $1-8\times 10^{-9}$ \\
2 &  $1-3\times 10^{-6}$& 0.7438 & $1-1\times 10^{-6}$\\
3 &  $1- 1\times 10^{-5}$ & 0.5984 & $1-5\times 10^{-6}$ \\
\bottomrule \bottomrule
\end{tabular}
\caption{Results of $R_a(D)$ [in Eq.~\eqref{eqn:R_aD}] and $R'_a(D)$ [in Eq.~\eqref{eqn:R-approx}] for the TFIC at the critical point $g=1$ and non-critical point $g=1.5$, for the ground state ($a=0$) and first three excited states ($a=1,2,3$), and with $N_A=N/2$. The chain length is $N=40$, and the truncation dimension is chosen as $D=30$.}
\label{tab:R-table}
\end{table}

\section{Conformal field theory formulation}
\label{sec:cft}

The benchmark results for the TFIC indicate that \textit{at criticality, the truncated set of ground-state Schmidt vectors provides an efficient basis for approximating low-lying excited states, whereas the accuracy diminishes away from criticality.} This phenomenon was first reported in Ref.~\cite{Chepiga2017}. However, to the best of our knowledge, its origin has not been systematically investigated. Here we present a CFT perspective that provides analytical insight into those numerical observations.

To start, we notice that the quantity $R_a(D)$ can be written as
\begin{equation}
    R_a(D)=\sum_{\alpha,\beta=0}^{D-1} (\tilde{\Lambda}_a)_{\alpha\beta} (\tilde{\Lambda}_a)^*_{\alpha\beta}=\tr(\Pi_D^{A} \tilde{\Lambda}_a \Pi_D ^{\bar{A}} \tilde{\Lambda}_a^{\dagger}),
\end{equation}
where we have introduced the projector onto the first $D$ left (right) Schmidt vectors $\Pi_D^{A}=\sum_{\alpha=0}^{D-1}|v_{\alpha}\rangle \langle v_{\alpha}|$ ($\Pi_D^{\bar{A}}=\sum_{\alpha=0}^{D-1}|w_{\alpha}\rangle \langle w_{\alpha}|$).
Introducing the matrix $M_a=\tilde{\Lambda}_a \tilde{\Lambda}_a^\dg$, we can define a modified quantity 
\begin{equation}
    \label{eqn:R-approx}R'_a(D)=\tr(\Pi_D^{A} M_a)=\sum_{\alpha=0}^{D-1} (M_{a})_{\alpha\alpha},
\end{equation}
i.e., the truncated sum of the diagonal of $M_a$.
Numerically, we observe that $R_a(D)$ can be accurately approximated by $R'_a(D)$ at the critical point,
\begin{equation}
 R_a(D) \approx R'_a(D),    
\end{equation}
as shown in Table~\ref{tab:R-table}. In fact, by Cauchy-Schwarz inequality and noting $R'_a(D)\leq 1$, the difference between $R$ and $R'$ can be bounded by $R'_a(D)-R_a(D)\leq \sqrt{1-\tr(\Pi_D^{\bar{A}}\tilde{\Lambda}_a^\dg \tilde{\Lambda}_a)}$. When $N_A=N/2$, the two halves of the system are symmetric, leading to $\tr(\Pi_D^{\bar{A}}\tilde{\Lambda}_a^\dg \tilde{\Lambda}_a)=R'_a(D)$, from which $R$ is lower bounded by $R_a(D)\geq R'_a(D)-\sqrt{1-R'_a(D)}$. Therefore, for a given bond dimension $D$, if $R'_a(D)$ is close to 1, one can conclude that $R_a(D)$ is also close to 1. 

The matrix element of $M_a$ has a clear physical interpretation: it is the reduced density matrix of the excited state in the ground-state Schmidt basis,
\begin{equation}
    (M_{a})_{\alpha\beta}=\langle v_\alpha|\rho_a^A|v_\beta\rangle, 
\end{equation}
where $|v_{\alpha}\rangle$ and $|v_\beta\rangle$ are ground-state Schmidt vectors on the subsystem $A$, and $\rho_a^A$ is the reduced density operator of the excited state $|\Psi_a\rangle$ after tracing out degrees of freedom on the complement $\bar{A}$, $\rho_a^A=\tr_{\bar{A}}|\Psi_a\rangle\langle\Psi_a|$. This physical interpretation allows for CFT evaluation of $(M_a)_{\alpha\beta}$. 

We will derive a CFT expression for $(M_{a})_{\alpha\beta}$~\cite{Alcaraz2011,Cardy2016,LiuYH2023}, from which it follows that $(M_{a})_{\alpha\alpha}$ decays with the increasing of $\alpha$. Since the normalization of $\rho_a^A$ ensures $\lim_{D\ra d^{N_A}} R'_a(D)=1$, this exponential decay directly implies that $R'_a(D)$ approaches unity already at a finite $D$. By the argument above, $R_a(D)$ therefore also approaches unity at a finite $D$. We will focus on the case where $|\Psi_a\rangle$ is a \emph{primary} state in CFT.

\subsection{CFT derivation for $(M_{a})_{\alpha\beta}$}

In this subsection, we derive a CFT expression for $(M_{a})_{\alpha\beta}$. For readers who are not interested in the technical details, they may skip the derivations and proceed directly to Eqs.~\eqref{eqn:M-CFT-expression} and \eqref{eqn:F-expression}, which are central results of this subsection. 

To start, we first write the path integral representation of $\rho_a^A$ and then convert to operator formalism. The operator formalism of $\rho_a^A$ allows us to express $(M_{a})_{\alpha\beta}$ conveniently. 

The path integral representation of $\rho_a^A$ can be obtained from that of $|\Psi_a\rangle$. Consider a CFT on a circle with circumference $L$~\footnote{For simplicity, we perform the CFT derivation with periodic boundary conditions. Changing to open boundary conditions is expected to modify only subleading terms and therefore does not affect the qualitative behavior.}. Using the state-operator correspondence, the state $|\Psi_a\rangle$ is prepared by inserting the associated primary operator $O_a(z)$ at the past infinity $z_{-\infty}=-i\infty$, where $z=\sigma+it$ is the complex coordinate. In terms of the path integral, 
\begin{equation}
    \langle X|\Psi_a\rangle\propto\int D\phi(\sigma,t) \, e^{-S[\phi]}O_a(z_{-\infty})\delta_{\phi (\sigma ,0^-),X(\sigma)},
\label{eqn:path-integral-psia}
\end{equation}
where the integration region is the lower half cylinder $t<0^-$ and $0\leq \sigma\leq L$. Here, $\phi(\sigma,t)$ denotes the local field whose Euclidean action is $S[\phi]$, $O_a$ is a functional of the local field $\phi$, and $X(\sigma)$ is the boundary field configuration of $\phi$ at $t=0$, with $\sigma\in[0,L]$. Similarly, its complex conjugation is represented as
\begin{equation}
    \langle\Psi_a|X\rangle\propto\int D\phi(\sigma,t) \, e^{-S[\phi]}O_{a^*}(z_{+\infty})\delta_{\phi (\sigma ,0^+),X(\sigma)},
\end{equation}
where the integration region is the upper half cylinder $t>0^+$ and $0\leq \sigma\leq L$, and $a^*$ denotes the charge conjugation of $a$. The matrix element of the pure-state density matrix $\rho_a = |\Psi_a\rangle\langle \Psi_a|$ between the field configurations $|X\rangle$ and $|X'\rangle$, is then obtained by combining these two path integrals together, yielding
\begin{align}
    &\langle X|\rho_a|X'\rangle \propto \int D \phi(\sigma,t) \,  \nonumber\\
    &\quad \times e^{-S[\phi]} O_a(z_{-\infty}) O_{a^*}(z_{+\infty})\, \delta_{\phi(\sigma,0^-),X(\sigma)} \delta_{\phi(\sigma,0^+),X'(\sigma)}
\end{align}
where the integration region is $(t<0^-)\cup(t> 0^+)$ and $0\leq \sigma\leq L$. 

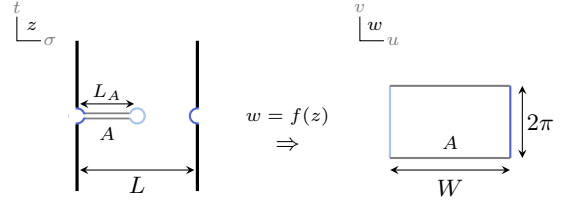
\begin{figure}
\begin{tikzpicture}[scale=.4,baseline={([yshift=-0.75ex] current bounding box.center)}]
\def\L{4};
\def\cl{0.8};
\def\cx{-2.};
\def\cxt{-1.};
\def\cy{2.5};
        \zslicenew{(0,0)};
        \draw (\cx, \cy+\cl) -- (\cx,\cy) -- (\cx+\cl,\cy);
        \draw (\cx+0.5,\cy+0.5) node {\scriptsize $z$};
        \draw[gray] (\cx+1.1,\cy) node {\scriptsize $\sigma$};
        \draw[gray] (\cx,\cy+1.1) node {\scriptsize $t$};
        \draw(\L*1.75,-1) node {$\Rightarrow$};
        \draw (\cxt+\L*2.6, \cy+\cl) -- (\cxt+\L*2.6,\cy) -- (\cxt+\cl+\L*2.6,\cy);
        \draw (\cxt+0.5+\L*2.6,\cy+0.5) node {\scriptsize $w$};
        \draw[gray] (\cxt+1.1+\L*2.6,\cy) node {\scriptsize $u$};
        \draw[gray] (\cxt+\L*2.6,\cy+1.1) node {\scriptsize $v$};
        \draw(\L*1.75,0) node {\scriptsize $w=f(z)$};
        \wslicenew{(\L*2.6,-0.2)}{1}{1};
    \end{tikzpicture}
    \caption{
    Integration region for the path integral representation of $\langle X_A|\rho_a^A|X'_A\rangle$, before and after the conformal transformation $w=f(z)$ from an infinite cylinder to a rectangle of height $2\pi$. For the $z$ coordinate, $\sigma=0$ and $\sigma=L$ are identified, thereby being a cylinder. The ultraviolet cutoffs $\epsilon$ at the entanglement points are represented by the blue curves and lines. In $\langle X_A|\rho_a^A|X'_A\rangle$, the boundary field configurations $X_A$ and $X'_A$ are imposed on the gray lines. }
    \label{fig:cfttrans}
\end{figure}

To examine the reduced density matrix,  we choose the subsystem $A=(\epsilon,L_A-\epsilon)$ and its complement $\bar{A}=(L_A+\epsilon,L-\epsilon)$, where $\epsilon$ is a short-distance cutoff at the entanglement points, with $\epsilon \ll L_A \leq L/2$.
Let us denote a field configuration on the subsystem $A$ as $X_A$ and on its complement $\bar{A}$ as $X_{\bar{A}}$. The matrix element of the reduced density matrix $\rho_a^A$ is obtained by tracing the full density matrix over the degrees of freedom in $\bar{A}$,
\begin{align}
    &\langle X_A|\rho_a^A|X'_A\rangle = \int D X_{\bar{A}} \,\langle X_A\cup X_{\bar{A}}|\rho_a^A|X'_A\cup X_{\bar{A}}\rangle \nonumber\\
    & \propto \int D \phi(\sigma,t) \, e^{-S[\phi]}\nonumber\\
    &\times O_a(z_{-\infty}) O_{a^*}(z_{+\infty})\, \delta_{\phi(\sigma\in A,0^+),X_A(\sigma)}\delta_{\phi(\sigma\in A,0^-),X'_{A}(\sigma)} \, .
    \label{eq:rhoA-z-coordinate}
\end{align}
This path integral is defined on a space-time geometry corresponding to an infinitely long cylinder with two holes of radius $\epsilon$, centered at $(0,0)$ and $(L_A,0)$, which is conformally equivalent to a sphere with two punctures. The region is shown schematically in Fig.~\ref{fig:cfttrans} (left). The two holes originate from the ultraviolet cutoff at the entanglement points, and the edges of the holes are imposed with specific conformal boundary conditions that encode the microscopic details of the regularization~\cite{Ohmori2015}.

To simplify the integration region, we use the following conformal transformation:
\begin{equation}
    w = f(z)=\log\left[-\frac{\sin(\frac{\pi(z-L_A)}{L})}{\sin (\frac{\pi z}{L})}\right],
\label{eq:conformal-transformation}
\end{equation}
where the branch of the logarithm is defined by $\log (z) := \log |z| + i\mathrm{Arg}(z)$, with $\mathrm{Arg}(z) \in (-\pi,\pi]$.
It maps the original spacetime with coordinate $z$, which is an infinite cylinder with two holes, to a rectangle parametrized by $w=u+iv$~\cite{Alcaraz2011}, where $u=(-W/2,W/2)$ and $v\in(-\pi,\pi]$; see Fig.~\ref{fig:cfttrans}. The width $W$ of the annulus is given by
\begin{align}
    W = 2|f(\epsilon)|= 2\log \left[ \frac{L}{\pi \epsilon}\sin (\frac{\pi L_A}{L})\right] .
\end{align}
The boundaries of the two original holes (at entanglement points) are mapped to the left and right edges of the rectangle, located at $u=\pm W/2$. The future and past infinity points $z_{\pm \infty} = \pm i \infty$ are mapped to $w_{\pm \infty}= \pm i\pi\frac{L_A}{L}=\pm i\pi r$.

In the new geometry ($w$-coordinate), the path integral in Eq.~\eqref{eq:rhoA-z-coordinate} can be interpreted as a boundary CFT (BCFT) on the rectangle. In this geometry, the matrix element of the reduced density matrix admits the following path integral representation~\cite{Alcaraz2011}:
\begin{align}
    &\langle X|\rho_a^A|X'\rangle 
    \propto \int D \phi(u,v) \, e^{-S[\phi]}\nonumber\\
    &\; \times O_a(w_{-\infty}) O_{a^*}(w_{+\infty})\, \delta_{\phi(u,\pi),X(u)} \delta_{\phi(u,-\pi),X'(u)}\, ,
    \label{eq:rhoA-w-coordinate}
\end{align}
where the integration region is over the rectangle, $u\in(-W/2,W/2)$ and $v\in(-\pi,\pi]$. We obtain the corresponding operator formalism by quantizing the theory along the $u$-direction (the rectangle space
direction). This yields
\begin{align}
    \rho_a^A \propto e^{-\pi \HBCFT} \hat{O}_{a^*}(w_{+\infty}) \hat{O}_{a}(w_{-\infty}) e^{-\pi \HBCFT}\, ,
\end{align}
where 
\begin{align}
\label{eqn:CFT-HE}
    \HBCFT = \frac{\pi}{W}\left(L_0 -\frac{c}{24}\right)
\end{align}
is the Hamiltonian for the BCFT on the rectangle, $\hat{O}_{a^*}(w_\infty)=e^{\pi r\HBCFT} \hat{O}_{a^*}(0) e^{-\pi r \HBCFT}$ is an operator in the Heisenberg picture, and similar for $\hat{O}_a(w_{-\infty})$. We note that $\HBCFT$, as the generator of translation in the $v$ direction, is identified with the entanglement Hamiltonian of $\rho_0^A$~\cite{Cardy2016}. The overall normalization factor is fixed by the condition $\mathrm{Tr}[\rho_a^A]=1$, leading to the final expression,
\begin{align}
    \rho_a^A = \frac{e^{-\pi \HBCFT} \hat{O}_{a^*}(w_{+\infty}) \hat{O}_{a}(w_{-\infty}) e^{-\pi \HBCFT}}{Z_{\mathrm{Anl}}\, \langle O_{a^*}(w_{+\infty}) O_{a}(w_{-\infty})\rangle_{\mathrm{Anl}}}\, ,
\end{align}
where ``Anl'' denotes the annulus region which is the previous rectangle region with $v=\pi$ and $v=-\pi$ identified; $\langle O_{a^*}(w_{+\infty}) O_{a}(w_{-\infty})\rangle_{\mathrm{Anl}}$ denotes the two-point correlator on the annulus, and $Z_{\mathrm{Anl}}$ is the annulus partition function $Z_{\mathrm{Anl}}=\int D\phi(u,v) e^{-S[\phi]}$.

Next, to evaluate $(M_a)_{\alpha\beta}$, we note that the Schmidt vectors $\{|v_\alpha\rangle\}$  are eigenstates of the entanglement Hamiltonian~\cite{Cardy2016}. For two eigenstates $|v_{\alpha}\rangle$ and $|v_{\beta}\rangle$ of the entanglement Hamiltonian $H$ in \eqref{eqn:CFT-HE} with $L_0$ eigenvalues $h_{\alpha}$ and $h_{\beta}$,  the matrix element of the reduced density operator can be written as
\begin{align}
\label{eqn:M-CFT-expression}
    (M_{a})_{\alpha\beta} = \langle v_\alpha|\rho_a^A|v_\beta\rangle = \frac{1}{Z_{\mathrm{Anl}}} e^{-\frac{\pi^2}{W}(h_\alpha + h_\beta -\frac{c}{12})} f^{a}_{\alpha\beta}(r)\, ,
\end{align}
where the function $f^{a}_{\alpha\beta}(r)$ is defined as
\begin{align}
\label{eqn:F-expression}
    f^{a}_{\alpha\beta}(r) := \frac{\langle v_\alpha| \hat{O}_{a^*}(w_{+\infty}) \hat{O}_{a}(w_{-\infty}) |v_\beta\rangle}{\langle O_{a^*}(w_{+\infty}) O_{a}(w_{-\infty})\rangle_{\mathrm{Anl}}}\, .
\end{align}
Equations~\eqref{eqn:M-CFT-expression} and \eqref{eqn:F-expression} are the central results of this section. The matrix element $ (M_{a})_{\alpha\beta}$ consists of a decaying factor $e^{-\frac{\pi^2}{W}(h_\alpha + h_\beta -\frac{c}{12})}$ along with a factor $f^{a}_{\alpha\beta}(r)$ which includes a correlator in BCFT.  
We remark that the above CFT derivation implicitly assumes that the conformal boundary conditions at the entanglement cuts are identical for the ground state and for the excited states associated with CFT primary fields. This is not known a priori and is therefore adopted as an assumption in our derivation. Under this assumption, the reduced density operator $\rho_a^A$ of a primary excited state $a$ is supported on the same Hilbert space as the ground-state reduced density operator $\rho_0^A$. This assumption is further supported by our numerical results for the critical Ising and three-state clock chains presented in Sec.~\ref{sec:transition}.

\subsection{Truncated basis approximation}

In this subsection, we show that $R'_a(D)$ defined in Eq.~\eqref{eqn:R-approx} is close to unity for low-lying $a$ (restricted to CFT primary states in this work) at a finite bond dimension $D\ll d^{N_A}$. In particular, for CFT primary states $a$, we show that the diagonal elements $(M_a)_{\alpha\alpha}$ decay exponentially with increasing $\alpha$. Given the normalization of $\rho_a^A$, which enforces $\lim_{D\ra d^{N_A}} R'_a(D)=1$, such rapid decay ensures that $R'_a(D)$ is close to unity at a finite $D$. 

As a consistency check, we first evaluate $(M_{0})_{\alpha\beta}$ for the ground state $a=0$, where $O_0=\bo$. This gives rise to $f^0_{\alpha\beta}(r)=\delta_{\alpha\beta}$ and
\begin{equation}
\label{eqn:M0-expression}
    (M_{0})_{\alpha\beta}=\frac{1}{Z_{\mathrm{Anl}}} e^{-\frac{2\pi^2}{W}(h_\alpha-\frac{c}{24})}\delta_{\alpha\beta}.
\end{equation}
As expected, the diagonal elements $(M_{0})_{\alpha\alpha}$ decay exponentially with the conformal weight $h_\alpha$. Since $\HBCFT$ is a BCFT Hamiltonian whose spectrum consists of equally spaced levels within each conformal tower, $h_\alpha$ grows approximately linearly with the index $\alpha$. Consequently, $(M_{0})_{\alpha\alpha}$ decays exponentially as $\alpha$ increases. 

Next, let us consider a low-lying excited state $a$ that corresponds to a CFT primary state. Since the Schmidt vectors $|v_\alpha\rangle$ and $|v_\beta\rangle$ are normalized, the coefficient $f_{\alpha\beta}^a(r)$ in Eq.~\eqref{eqn:M-CFT-expression} is (at most) of order $\mathcal{O}(1)$. It then follows that, to leading order, the matrix elements $(M_a)_{\alpha\beta}$ decay with the conformal weight $h_{\alpha}$ and $h_{\beta}$,
\begin{equation}
    (M_{a})_{\alpha\beta}\sim e^{-\frac{\pi^2}{W}h_\alpha}e^{-\frac{\pi^2}{W}h_\beta}.
    \label{eqn:Ma-expression-approx}
\end{equation}
This exponential suppression explains why $R'_a(D)$ approaches unity already at a finite bond dimension $D$.

Obtaining an explicit form of $f_{\alpha\beta}^a(r)$ is generally difficult for a generic BCFT. However, it can be computed perturbatively using the operator product expansion (OPE). Let us introduce $\tilde{r}:=2\pi r$, which is the distance between $w_{+\infty}$ and $w_{-\infty}$. The OPE takes the form
\begin{align}
    &\hat{O}_{a^*}(w_{+\infty}) \hat{O}_{a}(w_{-\infty})\nonumber\\
    &\quad = \frac{1}{\tilde{r}^{4h_a}} + \sum_{b\neq 0} \frac{1}{\tilde{r}^{4h_a-2h_b}} C_{aa^*}^b \hat{O}_b(0) + \cdots\, ,
\end{align}
where the sum runs over primary fields, and $C_{aa^*}^b$ are the corresponding OPE coefficients. This expansion yields the matrix elements
\begin{align}
    &\langle v_{\alpha}| \hat{O}_{a^*}(w_{+\infty}) \hat{O}_{a}(w_{-\infty}) |v_{\beta}\rangle \nonumber\\
    &\quad = \frac{1}{\tilde{r}^{4h_a}} \left[\delta_{\alpha\beta} + \sum_{b\neq 0}  C_{aa^*}^b C_{\alpha b\beta}\, \tilde{r}^{2h_b} + \cdots\right]\, ,
\end{align}
and the annulus two-point function
\begin{align}
    &\langle O_{a^*}(w_{+\infty}) O_{a}(w_{-\infty})\rangle_{\mathrm{Anl}}\nonumber\\
    &\quad = \frac{1}{\tilde{r}^{4h_a}} \left[1 + \sum_{b\neq 0}  C_{aa^*}^b \mathcal{A}_b\, \tilde{r}^{2h_b} + \cdots\right]\, ,
\end{align}
where $C_{\alpha b\beta} = \langle v_{\alpha} |\hat{O}_b(0)|v_{\beta}\rangle$ are BCFT three-point coefficients, and $\mathcal{A}_b = \langle O_{b}(0)\rangle_{\mathrm{Anl}}$ are one-point amplitudes. Combining the two results yields
\begin{align}
\label{eqn:f-expansion}
    f^a_{\alpha\beta}(r) = \delta_{\alpha\beta} + \sum_{b\neq 0} C_{aa^*}^b (C_{\alpha b\beta} -\mathcal{A}_b)\tilde{r}^{2h_b} + \cdots\, .
\end{align}

We now discuss two limiting regimes of $r$. For $r=L_A/L\ll 1$, the dominant contribution comes from the identity term $\delta_{\alpha\beta}$ in Eq.~\eqref{eqn:f-expansion}, which leads to 
\begin{equation}
\label{eqn:es-small-r}
    (M_{a})_{\alpha\beta}\approx \frac{1}{Z_{\mathrm{Anl}}} e^{-\frac{2\pi^2}{W}(h_\alpha-\frac{c}{24})}\delta_{\alpha\beta}.
\end{equation}
Consequently, in the limit $r=L_A/L\ll 1$, the entanglement spectrum of a low-lying excited (primary) state $a$ resembles that of the ground state. In contrast, when $r=1/2$, the contributions from the subleading terms in Eq.~\eqref{eqn:f-expansion} can exceed the diagonal term $\delta_{\alpha\beta}$. In this regime, off-diagonal matrix elements, $(M_{a})_{\alpha\beta}$ with $\alpha\neq\beta$, are generally non-vanishing. This signals the onset of an entanglement spectrum transition as $r$ is increased from $r\ll 1$ toward $r=1/2$. We analyze this transition in detail using two numerical examples in the next section.

\begin{figure*}[tb]
    \includegraphics[width=0.75\textwidth]{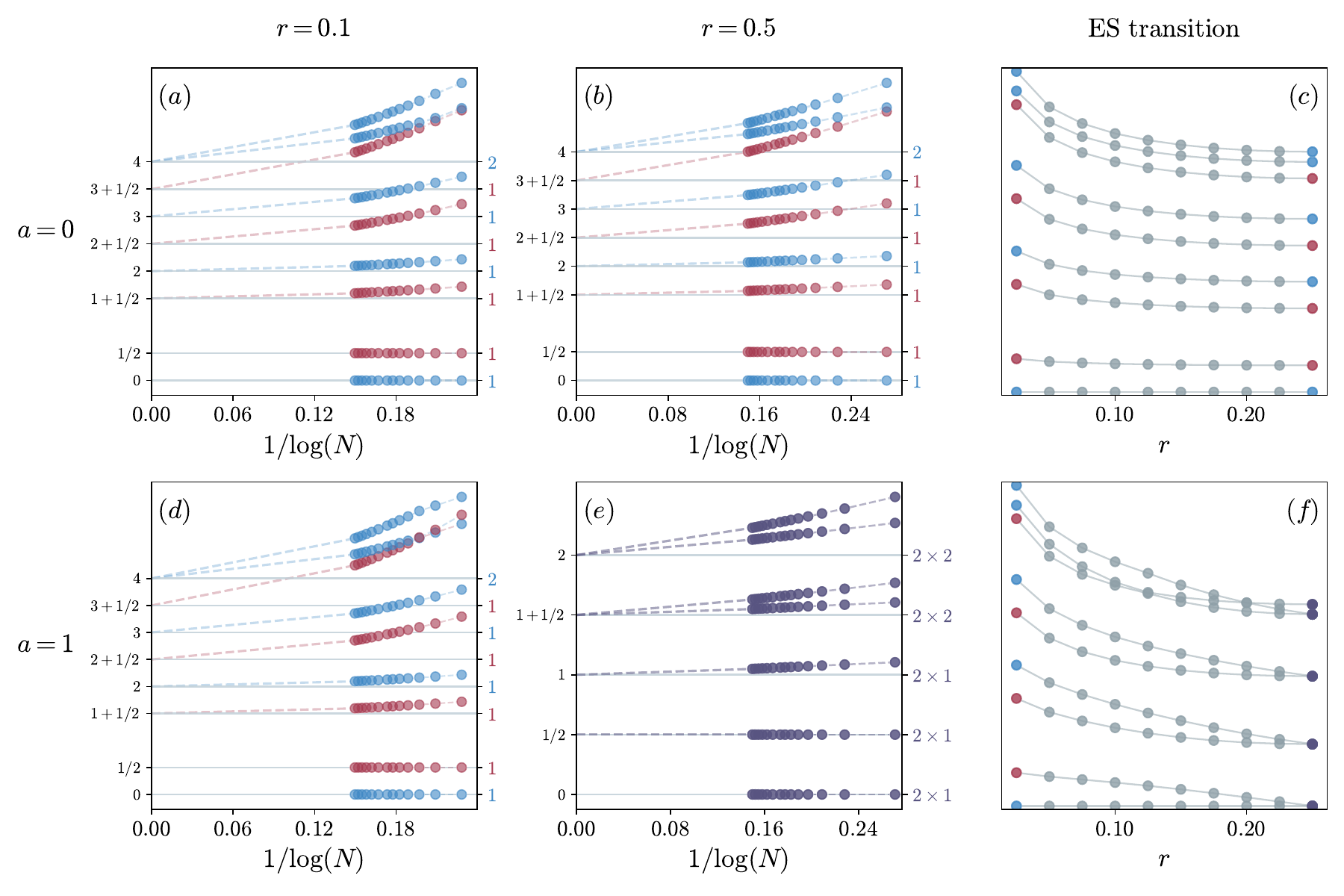}
    \caption{Entanglement spectrum $\tilde{\xi}$ of the critical Ising chain with open boundary conditions. (a) and (b) display the scaling of the ground state ($a=0$) entanglement spectrum with respect to system size $1/\log(N)$ for
    $N=40,80,120,160,200,240,280,320,400,480,560,640,720,800$, where the ratio $r=N_A/N$ is 0.1 for (a) and 0.5 for (b). For both (a) and (b), we shift and rescale the entanglement spectrum to $\tilde{\xi}$ such that the lowest level $\tilde{\xi}^{(a)}_0=0$ and the second-lowest level $\tilde{\xi}^{(a)}_1=1/2$. The vertical ticks on the left of the figure are the expected (or observed) conformal dimensions, where different expected conformal towers are plotted with different colors; the vertical ticks on the right are degeneracies in the thermodynamic limit. In the degeneracy label, ``$2\times 1$'' means the data exhibit exact two-fold degeneracy already at finite size. In (c), the system size is fixed at $N=800$, and we take $r\in[0.05,0.5]$ to illustrate the entanglement spectrum (ES) transition. We only shift the entanglement spectrum but do not rescale it. Both the vertical ticks are degeneracies. In (d), (e), and (f), we show the same analysis for the first excited state ($a=1$).}
    \label{fig:Ising-es}
\end{figure*}

\section{Entanglement spectrum transition}
\label{sec:transition}

For the lattice Hamiltonian at the critical point, it is well-known that the ground-state entanglement spectrum reflects the conformal tower structure associated with the conformal boundary condition~\cite{Laeuchli2013}. In contrast, for the excited states, the results in Sec.~\ref{sec:cft} indicate a transition in the entanglement spectrum as the parameter $r$ is varied. In this section, we present numerical evidence that explicitly illustrates such transitions. 

The calculation is performed as follows. Given the exact excited state $|\Psi_a\rangle$ (or the approximate stitched excited state $|\psi_a\rangle$) and its reduced density operator $\rho_a^A$ of subsystem $A$, 
the entanglement spectrum is obtained from the eigenvalues of $\rho_a^A$. Denoting by $\{s^{(a)}_{\alpha}\}_{\alpha=0,\ldots,D-1}$ the eigenvalues of $\rho_a^A$, the entanglement spectrum is defined by
\begin{equation}
\label{eqn:entanglement-spectrum}
    \xi^{(a)}_{\alpha}=-\frac{1}{2\pi}\log s^{(a)}_{\alpha}.
\end{equation}
Namely, $\{\xi^{(a)}_{\alpha}\}$ is the spectrum of entanglement Hamiltonian $H_{E}^{(a)}$ defined via $\rho_a^A = e^{-2\pi H_{E}^{(a)}}$. When $|\Psi_0\rangle$ is the ground state of a critical Hamiltonian, $H_E^{(0)}$ is expected to be a critical Hamiltonian with the open boundary condition whose continuum limit is $\HBCFT$ in Eq.~\eqref{eqn:CFT-HE}, and therefore one expects the entanglement spectrum $\{\xi^{(0)}_{\alpha}\}$ to be equally spaced and exhibit conformal tower structures. 

To reveal the scaling feature and conformal towers in the entanglement spectrum in the numerical calculations, we first shift the lowest level to zero and then rescale the second-lowest level to a fixed value $h$,
\begin{equation}
\begin{aligned}
    \tilde{\xi}^{(a)}_{\alpha} &= (\xi^{(a)}_{\alpha} - \xi^{(a)}_{0})\frac{h}{\xi^{(a)}_{1}-\xi^{(a)}_{0}}.
\end{aligned}
\end{equation}
We denote the spectrum after shifting and rescaling as $\tilde{\xi}^{(a)}_{\alpha}$. Using the result of Eq.~\eqref{eqn:M0-expression}, the ground-state entanglement spectrum in the continuum limit scales as $\xi^{(0)}_{\alpha}\sim \frac{1}{W}(h_\alpha-\frac{c}{24})$, where, for a fixed ratio $r$, $W\sim \log N$. We see that $\{\xi_\alpha^{(0)}\}$ is indeed proportional to the conformal towers $\{h_\alpha\}$. Furthermore, it shows that the length scale of $H_E^{(0)}$ is $\log N$, motivating us to plot the finite-size scaling of $\tilde{\xi}$ with respect to $1/\log N$ for a fixed ratio $r$. 

When the subsystem-to-system size ratio $r=N_A/N$ is small ($r \ll 1$), the entanglement spectrum of the excited states closely resembles that of the ground state to the leading order, as shown in Ref.~\cite{Alcaraz2011} and Eq.~\eqref{eqn:es-small-r}. In the opposite limit, $r=1/2$, the entanglement spectrum of the excited states deviates strongly from that of the ground state. Hence, as $r$ increases from $r \ll 1$ to 1/2, the entanglement spectrum undergoes a qualitative reorganization, which we refer to as an ``entanglement spectrum transition.''

We illustrate the entanglement spectrum transition using two models at their critical points: the TFIC [Eq.~\eqref{eqn:Ising-Hamil}] and the three-state clock model.

\subsection{Transverse-field Ising chain}

\begin{figure*}
    \centering
    \includegraphics[width=0.75\textwidth]{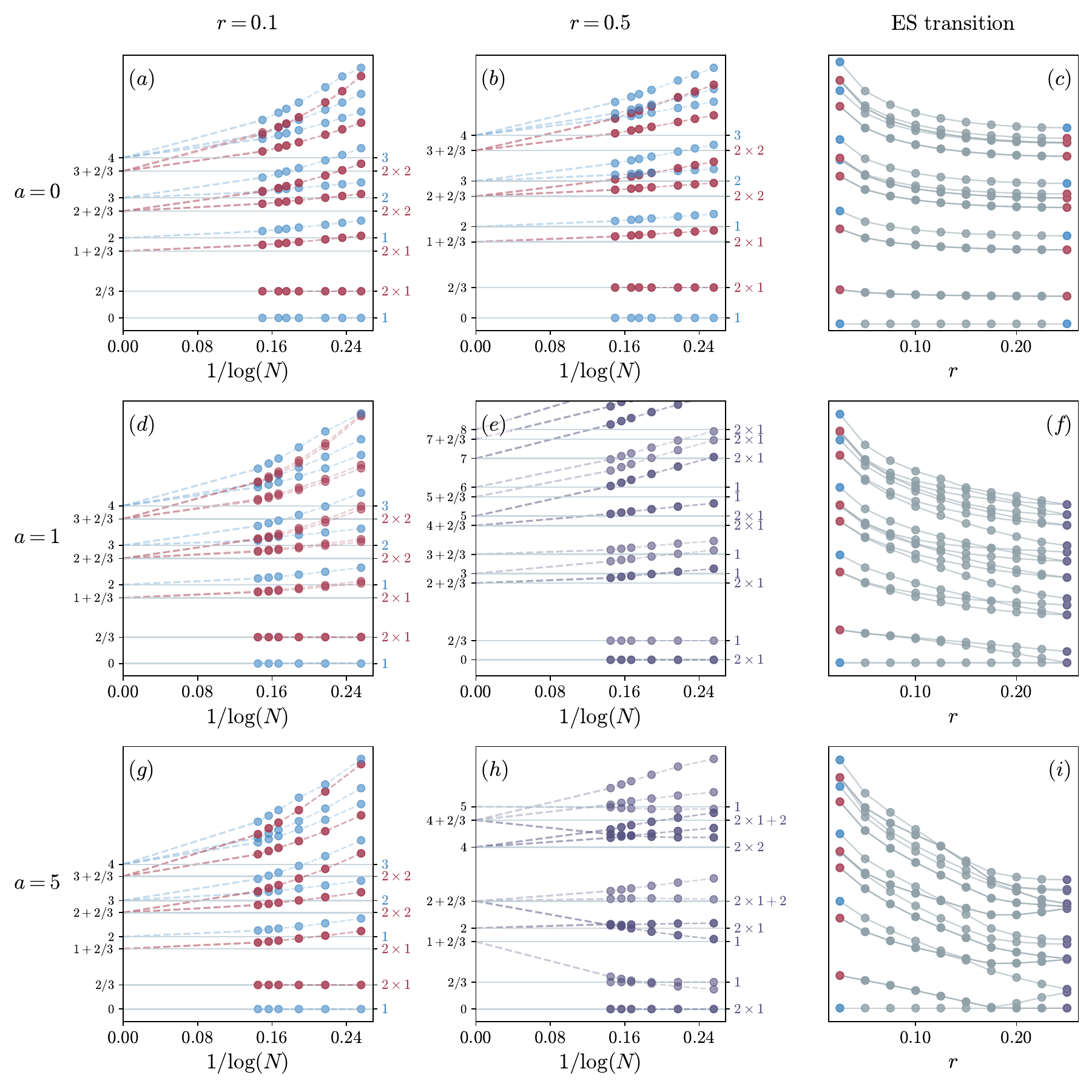}
    \caption{Entanglement spectrum $\tilde{\xi}$ for the critical three-state clock model with open boundary conditions. The system sizes are $50 \leq N \leq 1000$, and the bond dimension is up to $D=1000$. (a) and (b) display the scaling of the ground state ($a=0$) entanglement spectrum with respect to system size $1/\log(N)$ for where the ratio $r=N_A/N$ is 0.1 for (a) and 0.5 for (b). For both (a) and (b), we shift and rescale the entanglement spectrum to $\tilde{\xi}$ such that the lowest level $\tilde{\xi}^{(a)}_0=0$ and the second-lowest level $\tilde{\xi}^{(a)}_1=2/3$. The vertical ticks on the left of the figure are the expected (or observed) conformal dimensions, where different expected conformal towers are plotted with different colors; the vertical ticks on the right are degeneracies.  In (c), the system size is fixed at $N=1000$, and we take $r\in[0.05,0.5]$ to illustrate the entanglement spectrum (ES) transition. We only shift the entanglement spectrum but do not rescale it. Both the vertical ticks are degeneracies. In (d), (e), and (f), we show the same analysis for the first excited state $a=1$; and in (g),(h), and (i), we show the same analysis for the excited state $a=5$. }
    \label{fig:clock}
\end{figure*}

As a first example, we consider the TFIC introduced in Sec.~\ref{sec:excited-state} at its critical point. 
The lattice Hamiltonian for the TFIC is given in Eq.~\eqref{eqn:Ising-Hamil}. At the critical point $g=1$, the Hamiltonian is critical, which in the thermodynamics limit, is Ising CFT with central charge $c=1/2$, and with three (chiral) primary fields, denoted as $I,\sigma,\varepsilon$. When considering an open chain with free boundary conditions on both ends, the energy spectrum consists of conformal towers of the primary operators $I$ and $\varepsilon$~\cite{Cardy1989}, where $\varepsilon$ has conformal dimension $h_\varepsilon=1/2$. 

Using the mapping to a free-fermion model, we can express the exact eigenstate $|\Psi_a\rangle$ and its reduced density operator $\rho_a^A$, 
which amounts to choosing $D=2^{N_A}$ and allows efficient evaluation of the exact entanglement spectrum for large system sizes (see details in Appendix~\ref{app:free-fermion-es}). 
We show the numerical results for the entanglement spectrum in Fig.~\ref{fig:Ising-es} with different ratios $r$ and for different exact eigenstates $|\Psi_a\rangle$. 

For the ground state $a=0$, Figs.~\ref{fig:Ising-es}(a) and \ref{fig:Ising-es}(b) show that the entanglement spectrum of $r=0.1$ and $r=0.5$ both resemble the energy spectrum with the free boundary condition, as expected, with two conformal towers corresponding to those of $I$ and $\varepsilon$. There is no entanglement spectrum transition as shown in Fig.~\ref{fig:Ising-es}(c).

For the first excited state $a=1$ (corresponding to the primary state $|\varepsilon\rangle$), the entanglement spectra of $r=0.1$ and $r=0.5$ are different, as shown in Figs.~\ref{fig:Ising-es}(d) and \ref{fig:Ising-es}(e). For the entanglement spectrum at $r=0.5$, we shift and rescale such that $\tilde{\xi}^{(1)}_1=1/2$. At $r=0.5$, the entanglement spectrum is still equally spaced, albeit with a very different conformal tower structure. Interestingly, the entanglement spectrum at $r=0.5$ possesses an exact two-fold degeneracy at every level. Such degeneracy comes from a combination of the reflection symmetry and the $\mathbb{Z}_2$ symmetry $Q=\prod_i X_i$~\cite{You2022}. The entanglement spectrum transition manifests itself in Fig.~\ref{fig:Ising-es}(f). 

\subsection{Three-state clock model}

Moving beyond the TFIC, we next consider the 1D three-state clock model, which is a genuine interacting model. The lattice Hamiltonian 
is given by
\begin{equation}
    H_{\text{clock}} = -\sum_{i=1}^{N-1} ( Z_i Z_{i+1}^\dagger + Z_i^\dagger Z_{i+1} ) - g\sum_{i=1}^{N}(X_i + X_i^\dagger) ,
\end{equation}
where each site hosts 
a degree of freedom of dimension $d=3$. The operators $X_i$ and $Z_i$ can be represented as
\begin{equation}
Z_i = 
\begin{bmatrix}
1 & 0 & 0\\
0 & \omega & 0 \\
0 & 0 & \omega^2 \\
\end{bmatrix}, \quad
X_i = 
\begin{bmatrix}
0 & 1 & 0  \\
0 & 0 & 1  \\
1 & 0 & 0
\end{bmatrix}
\end{equation}
with $\omega = e^{\frac{2\pi i}{3}}$. The Hamiltonian $H_{\text{clock}}$ has a $\mathbb{Z}_3$ symmetry and commutes with the $\mathbb{Z}_3$ charge operator $Q=\prod_i X_i$. Hence eigenstates of $H_{\text{clock}}$ can be labeled by their $\mathbb{Z}_3$ charge $q$ via $Q |\Psi\rangle = \omega^q |\Psi\rangle \; (q=0,1,2)$. At the phase transition point $g=1$, the three-state clock model is critical and described by the $\mathbb{Z}_3$ parafermion CFT with central charge $c=4/5$~\cite{Zamolodchikov1985,LiW2014}. For an open chain with the free boundary condition on both ends, the energy spectrum consists of three conformal towers associated with the primary fields $I$, $\psi$, and $\psi^\dg$~\cite{Affleck1998}, where $\psi$ and $\psi^\dg$ have the same conformal dimension $h_\psi = h_{\psi^{\dagger}} =2/3$. 

Since the exact solution does not provide access to the entanglement spectrum of excited states, we take the excited states as $|\psi_{a}\rangle$ using the stitched excitation ansatz in Sec.~\ref{sec:excited-state}.
To this end, we use a variational algorithm to find an MPS approximation for the ground state of the model, for system sizes ranging from $N=50$ to $N=1000$ and open boundary conditions, using bond dimension up to $D=1000$.
The reduced density operator $\rho_a^A$ in the truncated Schmidt vectors basis $\{ |v_\alpha\rangle\}_{\alpha=0,\ldots, D-1}$ gives the matrix $M_a$.

The numerical results for the entanglement spectrum are shown in Fig.~\ref{fig:clock} for different ratios $r$ and different eigenstates $|\psi_a\rangle$. For the ground state $a=0$, Figs.~\ref{fig:clock}(a) and \ref{fig:clock}(b) shows that the entanglement spectrum of $r=0.1$ and $r=0.5$ both resemble the CFT spectrum with the free boundary condition, as expected, where three conformal towers manifest. 

Turning to the excited states, the entanglement spectrum exhibits drastic changes for different ratios $r$. For a small ratio $r=0.1$, the spectrum still resembles that of the ground state, displaying three clear conformal towers, as shown in Figs.~\ref{fig:clock}(d) and \ref{fig:clock}(g). In contrast, for $r=0.5$, the conformal towers reorganize but the spectrum is still equally spaced, as shown in Figs.~\ref{fig:clock}(e) and \ref{fig:clock}(h). In the energy spectrum,  the excited state $a=1$ and $a=2$ are degenerate, corresponding to primary states $|\psi\rangle$ and $|\psi^\dg\rangle$, respectively. To remove ambiguity in numerics, we note that $|\psi\rangle$ and $|\psi^\dg\rangle$ are distinguished by their $\mathbb{Z}_3$ charges, $Q|\psi\rangle = \omega|\psi\rangle$ and $Q|\psi^\dg\rangle = \omega^2|\psi^\dg\rangle$, allowing us to single out $|\psi\rangle$ from the degenerate subspace. For the entanglement spectrum of the excited state $a=1$ at $r=0.5$, we shift and rescale such that $\tilde{\xi}^{(1)}_1=2/3$. We observe that the resulting spectrum $\tilde{\xi}$ contains one tower with integer conformal dimensions, and another with conformal dimensions given by integers plus $2/3$. We also examine the excited state $a=5$, which is a descendant state in the conformal tower of $I$ and is non-degenerate. At $r=0.5$, we observe that the entanglement spectrum $\tilde{\xi}$ also consists of one tower with integer conformal dimensions, and another with conformal dimensions as integers plus $2/3$, but the degeneracies are different from those of $a=1$. Moreover, the finite-size effect in the entanglement spectrum of $a=5$ is also larger than that of $a=1$. Since $a=5$ state is a descendant state, our CFT analysis (targeting primary states) may not apply. Nevertheless, we still observe the entanglement spectrum transition numerically.

\section{Summary and outlook}
\label{sec:discussion}

In this work, we have provided a conformal field theory perspective that clarifies why the low-lying excited states of one-dimensional critical systems can be constructed from the local effective Hamiltonian obtained from the MPS approximation to the ground state. Our analysis supports the intuition that, at criticality, the truncated basis formed by the ground-state Schmidt vectors serves as an efficient basis for expanding excited states. By expressing the matrix elements of the excited-state reduced density operator in the basis of ground-state Schmidt vectors as CFT correlation functions, we have shown that the contributions of higher Schmidt vectors decay, thereby explaining the numerical success of this approach.

Our CFT derivation applies to the low-lying excited states that are primary states. In numerical simulations, however, low-lying descendant states are also accurately captured by the stitched-excitation ansatz, suggesting that the CFT argument could likely be extended to them as well. Deriving an explicit CFT expression for $\rho_a^A$ when $a$ is a descendant is more intricate, and we leave this for future investigation. It would also be interesting to explore broader settings, for example, thermal states~\cite{Cardy2014} and global quantum quenches~\cite{Calabrese2007}, where the path integral formalism is also applicable.

Furthermore, the CFT analysis predicts an entanglement spectrum transition as the ratio $r$ between the subsystem size and the total chain length is varied. Our numerical results support this prediction: as $r$ increases from $r\ll 1$ to $r=1/2$, the conformal towers in the entanglement spectrum reorganize. Identifying the nature of these conformal towers at $r=1/2$ remains an interesting open question. In particular, it is not known a priori whether they correspond to the conformal towers of (subsets of) the primary operators in the periodic chain; indeed, our numerical results suggest that they do not. In principle, Eq.~\eqref{eqn:M-CFT-expression} provides an analytical route to the reduced density matrix. However, evaluating the BCFT correlator in Eq.~\eqref{eqn:F-expression} requires additional analytical techniques and remains challenging, even for Virasoro minimal models, due to the involvement of descendant states and their intricate conformal block structure. Developing a theoretical understanding of the reorganization of the entanglement spectrum is an intriguing direction for future work.

\section*{Acknowledgments}
We thank Natalia Chepiga, Adri\'an Franco-Rubio, Germ\'an Sierra, Tao Xiang for insightful discussions. D.C. and M.C.B. acknowledge financial support by the Deutsche Forschungsgemeinschaft (DFG, German Research Foundation) under Germany's Excellence Strategy -- EXC-2111 -- 390814868; and Research Unit FOR 5522 (grant nr. 499180199), and by the EU-QUANTERA project TNiSQ (BA 6059/1-1). M.Y. is supported by the Distinguished Postdoc Fellowship of the Munich Center for Quantum Science and Technology (MCQST) funded by DFG under the Excellence Strategy EXC2111-390814868. Y.S.Z. is supported by the Sino-German (CSC-DAAD) Postdoc Scholarship Program. Y.L. is supported by the Alexander von Humboldt Foundation.

Raw data are available on Zenodo~\cite{cocchiarella2025excited}.

\bibliography{main}

\onecolumngrid

\appendix

\section{Free fermion technique}
\label{app:Gaussian}

\subsection{Wavefunction overlap}
In this appendix, we explain how to compute $R_a(D)$ in Eq.~\eqref{eqn:R_aD} for the transverse-field Ising model using the free fermion technique, specifically, the covariance matrix method. The quantity of interest is the overlap $(\langle v_{\alpha}| \otimes \langle w_{\beta}| ) | \Psi_a \rangle$ between an eigenstate $|\Psi_a\rangle$ and ground-state Schmidt vectors $|v_\alpha\rangle\otimes|w_\beta\rangle$. To this end, we consider the overlap of three general wavefunctions $|\phi_1\rangle,|\phi_2\rangle,|\phi_3\rangle$,
\begin{equation}
    (\langle\phi_1| \otimes \langle \phi_2|) |\phi_3\rangle ,
    \label{eqn:overlap-1}
\end{equation}
where $|\phi_3\rangle$ is defined on the full system, and $|\phi_1\rangle$ and $|\phi_2\rangle$ are supported on the subsystem $A$ and the complement $\bar{A}$, respectively. Diagrammatically, the overlap is represented as
\begin{equation}
    \begin{array}{c}
    \def\sft{0.5}
        \begin{tikzpicture}[scale=.4,baseline={([yshift=0ex] current bounding box.center)}]
        \bigtensor{(0,0)}{1}{8};
        \bigtensor{(0,\singledx*1.2)}{2}{4};
        \bigtensor{(\singledx*4,\singledx*1.2)}{2}{4};
        \draw(\singledx*2.5,\singledx*1.2) node {\scriptsize $\langle\phi_1|$};
        \draw(\singledx*4.5,0) node {\scriptsize $|\phi_3\rangle$};
        \draw(\singledx*6.5,\singledx*1.2) node {\scriptsize $\langle\phi_2|$};
        \end{tikzpicture}
        \end{array}.
\end{equation}

Suppose $|\phi_1\rangle,|\phi_2\rangle,|\phi_3\rangle$ are fermionic Gaussian states, the above overlap can be solved using the covariance matrix method. For a fermionic Gaussian state with density operator $\rho$, all information is contained in the covariance matrix $\Gamma$ encoding two-point correlators of Majorana operators~\cite{Bravyi2005},
\begin{equation}
    \Gamma_{jk} = \frac{i}{2} \tr(\rho [c_j, c_k]), 
\end{equation}
where Majorana operators satisfy $\{c_j,c_k\} = 2\delta_{jk}$. $\Gamma$ is a real and antisymmetric matrix satisfying $\Gamma^T \Gamma \leq \mathbbm{1}$, where $\Gamma^T \Gamma = \mathbbm{1}$ if and only if $\rho$ is a pure state. Using the covariance matrices, the norm of the overlap in Eq.~\eqref{eqn:overlap-1} can be calculated using a determinant formula~\cite{bravyi2017}:
\begin{equation}
\label{eqn:overlap-formula}
    |(\langle\phi_1| \otimes \langle \phi_2|) |\phi_3\rangle | = \left[ \det\left(\frac{\bo_{2N}-\Gamma^{|\phi_3\rangle}(\Gamma^{|\phi_1\rangle}\oplus\Gamma^{|\phi_2\rangle})}{2}\right) \right]^{\frac{1}{4}}
\end{equation}
with $\Gamma^{|\phi_i\rangle}$ being the covariance matrix of $|\phi_i\rangle$.

We now describe how to obtain the covariance matrices corresponding to the eigenstate $|\Psi_a\rangle$ and ground-state Schmidt vectors $|v_\alpha\rangle,|w_\beta\rangle$ for a quadratic Majorana Hamiltonian,
\begin{equation}
    H = \frac{i}{4}\sum_{j,k=1}^{2N} h_{jk} c_j c_k ,
\end{equation}
where $h$ is a $2N \times 2N$ real antisymmetric matrix. In fact, $H_{\text{Ising}}$ in Eq.~\eqref{eqn:Ising-Hamil} can be cast into this quadratic Majorana form after the Jordan-Wigner transformation,
\begin{equation}
    H_{\mathrm{Ising}}=-i\sum_{i=1}^{N-1} c_{2i}c_{2i+1}-i g\sum_{i=1}^{N} c_{2i-1} c_{2i}.
\end{equation}

First, we derive the covariance matrix for an eigenstate $|\Psi_a\rangle$ of $H$. The real antisymmetric matrix $h$ can be brought into a canonical form by an orthogonal matrix $Q$~\cite{Kitaev2001}, 
\begin{align}
h = Q^T \left[ \bigoplus_{m=1}^N 
\begin{pmatrix}
0 & -\varepsilon_m \\
\varepsilon_m & 0
\end{pmatrix}
 \right] Q
\end{align}
with $\varepsilon_m\geq 0 \; \forall m$. The orthogonal transformation defines a new Majorana basis, $c'_{m}=\sum_j Q_{mj} c_j$, under which the Hamiltonian becomes $H=-\frac{i}{2}\sum_{m=1}^N \varepsilon_m c'_{2m-1} c'_{2m}$, where all terms mutually commute and each term $ic'_{2m-1} c'_{2m}$ has eigenvalues $\pm 1$. The ground state $|\Psi_0\rangle$ satisfies $ic'_{2m-1} c'_{2m}|\Psi_0\rangle = |\Psi_0\rangle  \; \forall m$, from which one can derive its covariance matrix
\begin{equation}
    \Gamma^{|\Psi_0\rangle} = Q^T (\bo_N\otimes iY) Q,
\end{equation}
where $Y$ is the Pauli matrix. Generally, an eigenstate $|\Psi_a\rangle$ satisfies $ic'_{2m-1} c'_{2m}|\Psi_a\rangle = \sgn_a(m) |\Psi_a\rangle$ with $\sgn_a(m) = \pm 1$. The covariance matrix of $|\Psi_a\rangle$ in the original Majorana basis then reads
\begin{equation}
    \Gamma^{|\Psi_a\rangle} = Q^T \left[ \bigoplus_{m=1}^N 
\begin{pmatrix}
0 & \sgn_a(m) \\
-\sgn_a(m) & 0
\end{pmatrix} \right]
Q.
\end{equation}

Next, we derive the covariance matrices for the ground-state Schmidt vectors. We take the Schmidt vectors for the subsystem $A$ as an example. The reduced density operator of the subsystem $A$, defined as $\rho_0^A = \tr_{\bar{A}}|\Psi_0\rangle\langle\Psi_0|$, is a mixed fermionic Gaussian state. It covariance matrix is simply the $2N_A\times 2N_A$ upper-left submatrix of $\Gamma^{|\Psi_0\rangle}$~\cite{Peschel2003,Fidkowski2010}, which we denote as $\Gamma^{\rho^A_0}$. Similarly, $\Gamma^{\rho^A_0}$ can be brought into the canonical form by a $2N_A \times 2N_A$ orthogonal matrix $O$:
\begin{equation}
\Gamma^{\rho^A_0} = O^T \left[ \bigoplus_{m=1}^{N_A} 
\begin{pmatrix}
0 & \gamma_m \\
-\gamma_m & 0
\end{pmatrix} \right]
O
\end{equation}
with $0 \leq \gamma_m \leq 1 \; \forall m$. Since $\rho_0^A$ is still a fermionic Gaussian state, it is completely determined by its covariance matrix $\Gamma^{\rho^A_0}$. Switching to the new Majorana basis $\tilde{c}_m=\sum_{j=1}^{2N_A}O_{mj} c_j$, $\rho_{A}$ can be written as
\begin{equation}
\label{eqn:decompose-rho0}
\begin{aligned}
    \rho_0^A &= \prod_{m=1}^{N_A} \frac{1}{2}\left(1+i\gamma_m \tilde{c}_{2m-1} \tilde{c}_{2m}\right).
\end{aligned}
\end{equation}
Thus, the dominant Schmidt vector $|v_0\rangle$ corresponding to the largest eigenvalue of $\rho_0$ satisfies $i \tilde{c}_{2m-1} \tilde{c}_{2m} |v_0\rangle = |v_0\rangle \; \forall m$, whose covariance matrix reads
\begin{equation}
    \Gamma^{|v_0\rangle}=O^T(\bo_{N_A}\otimes iY)O.
\end{equation}
Similarly, the general Schmidt vector $|v_\alpha\rangle$ satisfies $i \tilde{c}_{2m-1} \tilde{c}_{2m} |v_\alpha\rangle = \sgn_{\alpha}(m) |v_\alpha\rangle$ with $\sgn_{\alpha}(m) = \pm 1$, and its covariance matrix is given by
\begin{equation}
    \Gamma^{|v_\alpha\rangle} = O^T \left[ \bigoplus_{m=1}^{N_A} 
\begin{pmatrix}
0 & \sgn_{\alpha}(m) \\
-\sgn_{\alpha}(m) & 0
\end{pmatrix} \right] O.
\end{equation}

\begin{figure}
\centering
\includegraphics[width=0.3\linewidth]{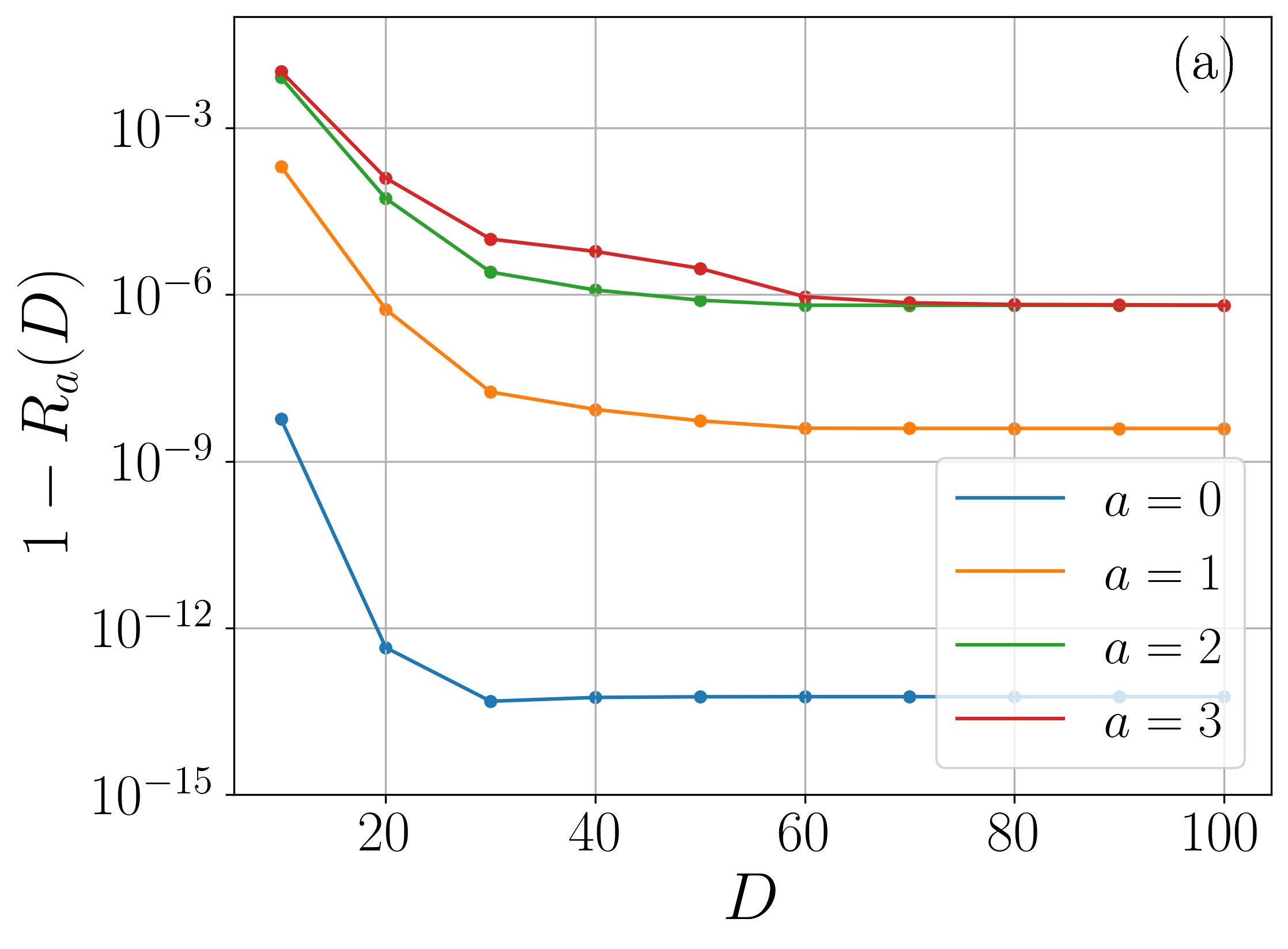}
\includegraphics[width=0.3\linewidth]{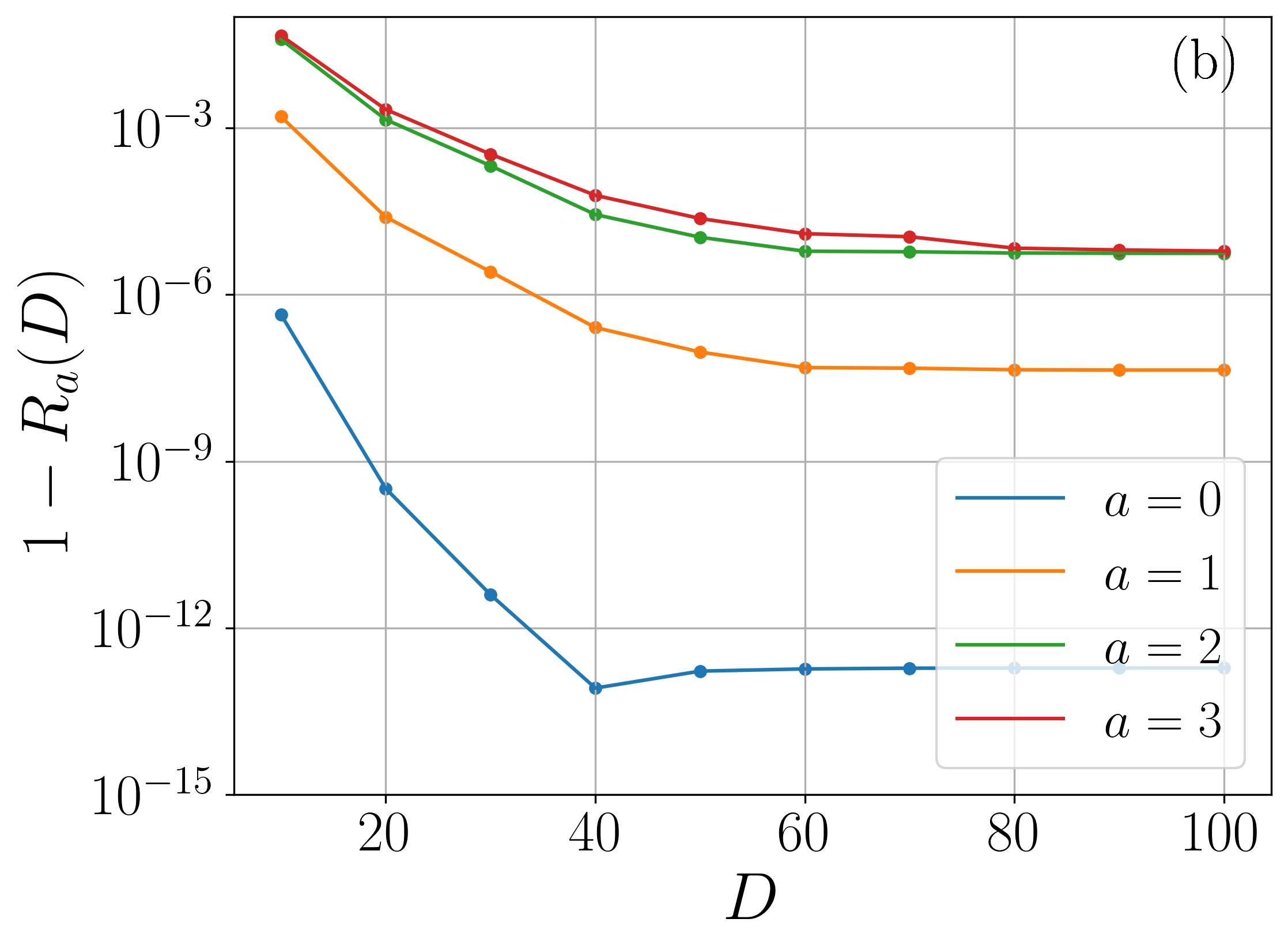}
\caption{Changes of $1-R_a(D)$ with respect to $D$, for the four lowest energy states of TFIC at the critical point $g=1$, with system size $N=40$ for (a) and $N=120$ for (b). }
\label{fig:R-with-D}
\end{figure}

The covariance matrix for $|w_\beta\rangle$ can be obtained in a similar way and will not be repeated. Once the covariance matrices $\Gamma^{|\Psi_a\rangle},\Gamma^{|v_\alpha\rangle},\Gamma^{|w_\beta\rangle}$ are obtained, one can compute the norm of the overlap $(\langle v_{\alpha}| \otimes \langle w_{\beta}| ) | \Psi_a \rangle$ using Eq.~\eqref{eqn:overlap-formula}. We show how $R_a(D)$ changes with $D$ in Fig.~\ref{fig:R-with-D}.

We remark that at the non-critical point of TFIM, for example, $g=1.5$, almost all singular values $\{\gamma_m\}$ of $\Gamma^{\rho_0^A}$ are extremely close to 1 up to the machine precision of the eigensolver, except for the first few $m$. As a result, the ordering of the higher Schmidt vectors $|v_\alpha\rangle$ can vary across different machines or numerical environments. One may observe differences in the value of $R_a(D)$ at $g=1.5$ in Table.~\ref{tab:R-table} when the computation is performed on different machines. 

Similarly, one can compute $R'_a(D)$ in Eq.~\eqref{eqn:R-approx} using the covariance matrix method, where the quantity of interest is the overlap $\langle v_\alpha|\rho_a^A|v_\alpha\rangle=\tr[\rho_a^A(|v_\alpha\rangle\langle v_\alpha|)]$, which is an overlap between two density matrices. Given two fermionic Gaussian states, represented by density operators $\omega_1,\omega_2$ and their corresponding covariance matrices $\Gamma^{\omega_1},\Gamma^{\omega_2}$, their overlap can be computed by
\begin{equation}
    \tr[\omega_1 \omega_2] = \sqrt{\det\left(\frac{\bo-\Gamma^{\omega_1}\Gamma^{\omega_2}}{2}\right)}. 
\end{equation}
Using the covariance matrices $\Gamma^{|v_\alpha\rangle}$ and $\Gamma^{\rho_a^A}$, one can then compute $\langle v_\alpha|\rho_a^A|v_\alpha\rangle$ and $R'_a(D)$.

\subsection{Entanglement spectrum}
\label{app:free-fermion-es}
In this section, we explain how to compute the entanglement spectrum in Eq.~\eqref{eqn:entanglement-spectrum} for the transverse-field Ising model using the covariance matrix method.

As an example, we show how to obtain $\{s^{(0)}_\alpha\}_{\alpha=0,1,\cdots,2^{N_A}-1}$ from $\Gamma^{\rho_0^A}$. From Eq.~\eqref{eqn:decompose-rho0}, $\rho_0^A = \prod_{m=1}^{N_A} \frac{1}{2}\left(1+i\gamma_m \tilde{c}_{2m-1} \tilde{c}_{2m}\right)$ and a general eigenstate $|v_\alpha\rangle$ of $\rho_0^A$ satisfy $i\tilde{c}_{2m-1}\tilde{c}_{2m}|v_\alpha\rangle=\sgn_\alpha(m)|v_\alpha\rangle$ with $\sgn_\alpha(m)=\pm 1$. Therefore, $\rho_0^A|v_\alpha\rangle=s_\alpha^{(0)}|v_\alpha\rangle$ with
\begin{equation}
    s_\alpha^{(0)}=\prod_{m=1}^{N_A} \frac{1}{2}(1+\sgn_\alpha(m) \gamma_m). 
\end{equation}
$\{s^{(a)}_\alpha\}$ can be obtained from $\Gamma^{\rho_a^A}$ using exactly the same method.

\end{document}